\documentclass[usenatbib,useAMS]{mn2e}

\usepackage{epsf}
\usepackage[dvips]{epsfig}
\usepackage{times}

\def\spose#1{\hbox to 0pt{#1\hss}}
\def\lta{\mathrel{\spose{\lower 3pt\hbox{$\mathchar"218$}}
     \raise 2.0pt\hbox{$\mathchar"13C$}}}
\def\gta{\mathrel{\spose{\lower 3pt\hbox{$\mathchar"218$}}
     \raise 2.0pt\hbox{$\mathchar"13E$}}}

\begin{document}

\label{firstpage}

\title[The Tidal Tails of NGC 5466]{The Tidal Tails of NGC 5466}

\author[M. Fellhauer et al.]{M. Fellhauer$^1$ \thanks{Email: madf, nwe,
vasily@ast.cam.ac.uk}, N.W. Evans$^1$, V. Belokurov$^1$,
M.I. Wilkinson$^2$ and G. Gilmore$^1$ \\ 
$^1$ Institute of Astronomy, University of Cambridge, Madingley Road,
Cambridge CB3 0HA, UK \\
$^2$ Dept.\ of Physics and Astronomy, University of Leicester,
University Road, Leicester LE1 7RH, UK}

\pagerange{\pageref{firstpage}--\pageref{lastpage}} \pubyear{2006}

\maketitle

\begin{abstract}
  The study of substructure in the stellar halo of the Milky Way has
  made a lot of progress in recent years, especially with the advent
  of surveys like the Sloan Digital Sky Survey.  Here, we study the
  newly discovered tidal tails of the Galactic globular cluster
  NGC~5466.  By means of numerical simulations, we reproduce the
  shape, direction and surface density of the tidal tails, as well as
  the structural and kinematical properties of the present-day
  NGC~5466.  Although its tails are very extended in SDSS data $(\gta
  45^\circ$), NGC~5466 is only losing mass slowly at the present epoch
  and so can survive for probably a further Hubble time.  The tidal
  shaping through the Milky Way potential, especially the potential
  of the disc, is the dominant process in the slow dissolution of
  NGC~5466 accounting for $\gta 60 \%$ of the mass loss over the
  course of its evolution.  The morphology of the tails provides a
  constraint on the proper motion -- the observationally determined
  proper motion has to be refined (within the stated error-margin) to
  match the location of the tidal tails. 
\end{abstract}


\section{Introduction}
\label{sec:intro}

Within the last few years, it has become more and more obvious that
the Milky Way stellar halo is dominated by substructure, particularly
dwarf galaxies, clouds, and tidal tails.  Data from the Sloan Digital
Sky Survey \citep[SDSS;][]{Yo00} have revealed abundant examples of
streams and substructure.  For example, \citet{Be06b} used a simple
colour cut $g-r < 0.4$ to map out the distribution of stars in SDSS
Data Release~5 (DR5).  The ``Field of Streams'', an RGB-composite
image composed of magnitude slices of the stellar density of these
stars, showed the overlap of the leading and trailing arm of the
well-known Sagittarius stream and the Monoceros ring very clearly.
Also prominent was a new stream, which did not have an identified
progenitor, and was called the ``Orphan Stream'' by~\citet{Be06b}.
The observational data on the Orphan Stream~\citep{Be07} was used by 
\citet{fe07} to argue that its progenitor may be the newly-discovered
disrupting dwarf galaxy UMa~II~\citep{Zu06}.

Tidal tails have proved to be an important diagnostic of the Galactic
potential.  Especially the tails of the dissolving Sagittarius dwarf
galaxy \citep*[see e.g.][]{Ib94,Ma03,He04,Jo05}, which wrap around the
Milky Way, are an excellent tracer of the strength and shape of the
potential.  \citet{fe06} have shown with their numerical
models that the bifurcation of the Sagittarius stream as seen in the
``Field of Streams'' is composed of two wraps of the tidal tails and
can only be reproduced if the orbital precession is small, i.e.\ if
the Milky Way dark matter halo is close to spherical.

Extra-tidal extensions and onsets of tidal tails have been claimed
around a number of Galactic globular clusters in recent years
\citep*[see e.g.][]{me01}.  The most spectacular and convincing
discovery remains the long and thin tail from the disrupting globular
cluster Pal~5 \citep{Od01,Ro02,Od03}.  The tails extend at least
$4$~kpc from the cluster in the leading and trailing direction and
contain more mass than the remaining cluster.

Recently, two different groups~\citep{Be06a,gr06} claim to have
detected tidal tails of various extents around the disrupting globular
cluster NGC 5466.  This is an old, metal-poor ([Fe/H]$ =-2.22$)
cluster, lying at Galactic coordinates $l=42.\!\!^o15$,
$b=73.\!\!^o59$.  In \citet{Be06a}, the observed tails of NGC~5466 are
not as long as those of Pal~5, stretching about $2^\circ$ or $500$~pc
in either direction.  \citet{gr06} reported afterwards that they found
evidence for a much larger extension of the tidal tails of NGC~5466.
They claimed that the leading arm extends over $\sim 30$~degrees and
the trailing arm extends at least $15$~degrees, before it leaves the
area covered by SDSS.  This finding makes the tails of NGC~5466 even
longer, but much fainter, than the tails of Pal~5.  The aim of our
paper is to confront these claims with theoretical expectation, as
well as to study the survival of the tails.

The following data for NGC~5466 are taken from various sources in the
literature \citep*{Ha96,Di99,Le97,Pr91}.  The central surface
brightness is $21.28$~mag\,arcsec$^{-2}$.  The total luminosity is
$M_{V}=-6.96$~mag and the mass-to-light ratio as given by \citet{Pr91}
is $\sim 1$.  Using these values, we derive a total mass of about $5
\times 10^{4}$~M$_{\odot}$.  The core radius has values ranging from
$6.1$ to $7.6$~pc; the half-mass radius ranges from $10.4$ to
$13.1$~pc.  The most substantial differences in the literature occur
for the tidal radius.  Here, values are spread between $61$ and
$158$~pc.  We use this data as constraints on our numerical
simulations.  Our aim is to access possible initial models of this
globular cluster and analyse its stability and evolution in different
sets of Milky Way potentials.

In the next section, we describe the setup of our simulations --
namely, the choice of Galactic potential models, the orbit of NGC~5466
and finally the initial model of the cluster itself.  This is followed
by a study of the relative importance of two-body relaxation and disc
shocking in Section~\ref{sec:sbox}, justifying our use of
particle-mesh simulations in this paper.  Then, in
Section~\ref{sec:res}, we present simulations that reproduce the
shape, extent and surface density of the tidal tails detected by
\citet{Be06a} and \citet{gr06}.  The properties of the remnant and
shown to correspond to the present-day NGC~5466.  Finally, we examine
how the tidal tails change as a function of the proper motion and
hence orbit.

\section{Setup}
\label{sec:setup}

\subsection{Galactic Models}
\label{sec:galpot}

\begin{figure}
\begin{center}
\includegraphics[height=4cm]{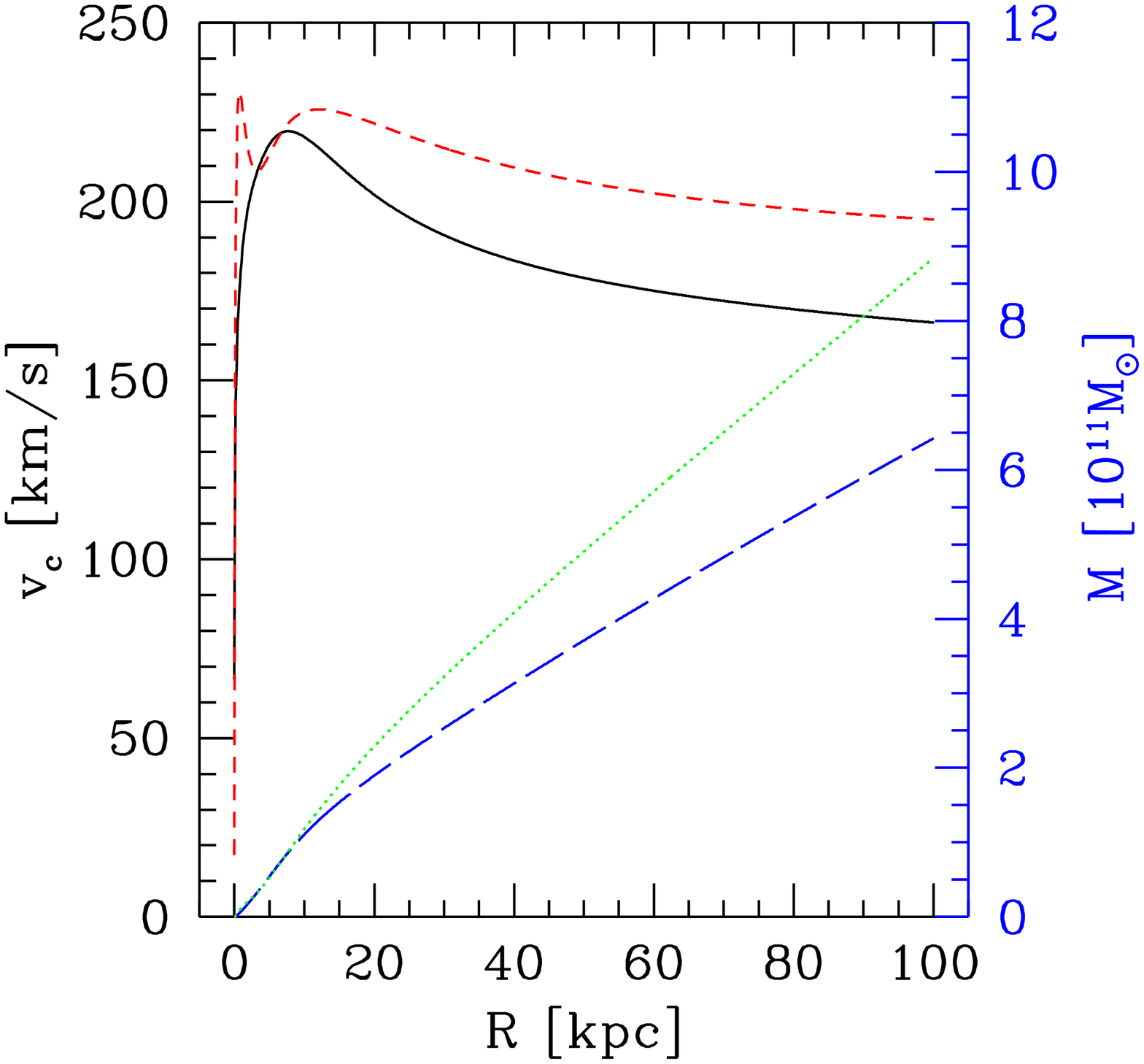}
\includegraphics[height=4cm]{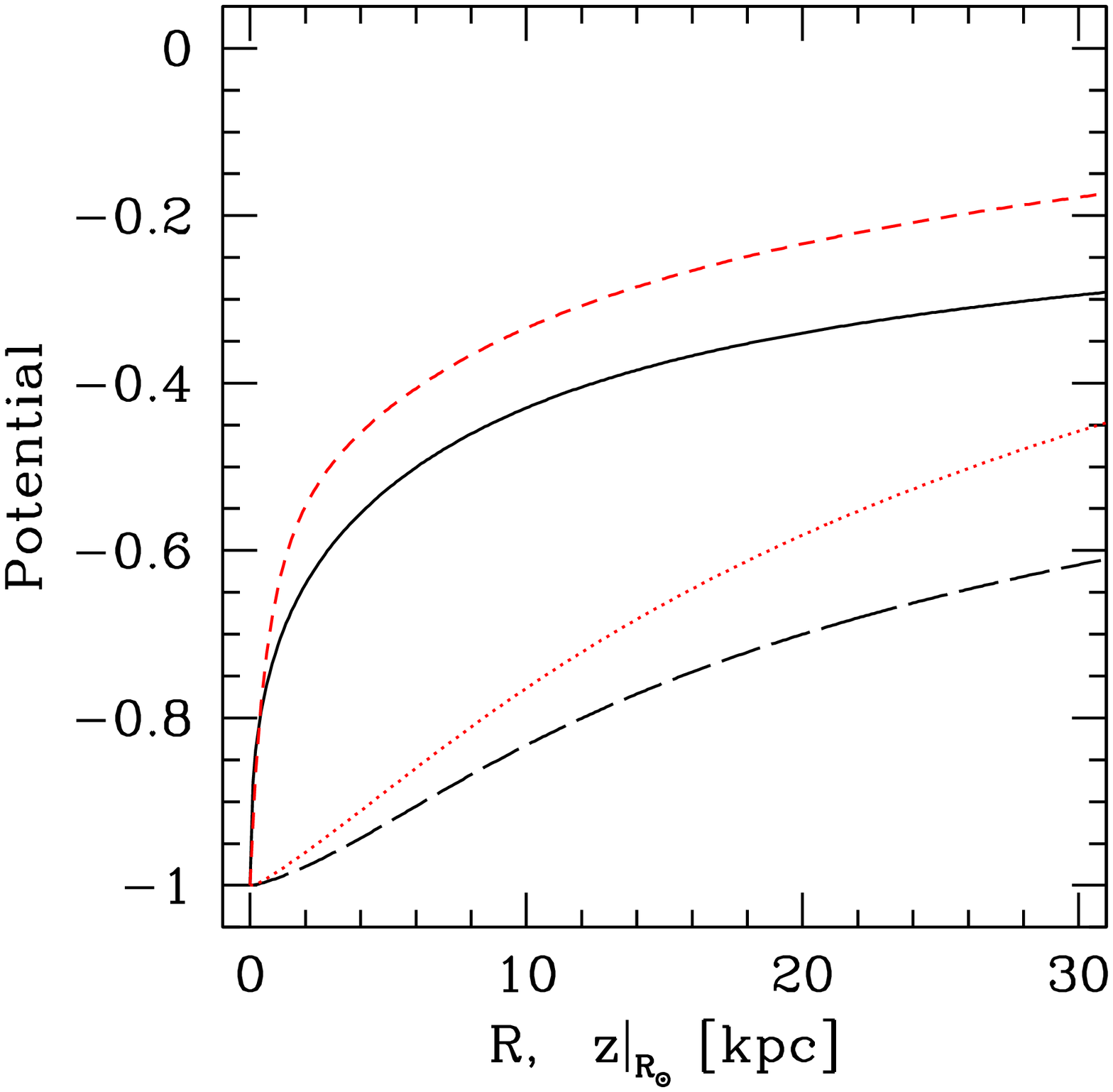}
\end{center}
  \caption{Comparison of the Galactic potentials.  Left: Circular
    velocity of the ML (solid, black) and DB (dashed, red) potentials.
    Also plotted is the enclosed mass (right hand $y$-axis) of the ML
    (dotted, green) and DB (long-dashed, blue) models.  Both
    potentials have the same circular velocity ($220$~km\,s$^{-1}$) at
    the solar radius.  Right: Gravitational potential in the disc
    plane for the ML (dashed) and DB (solid) models.  Also shown is
    the gravitational potential along the $z$-axis at the solar radius
    for the ML (dotted) and DB (long-dashed) models.  Even though the
    potentials agree well in the innermost part, the ML potential is
    much steeper in the outer parts.}
  \label{fig:potcomp}
\end{figure}

Dynamical friction does not play a significant role in the evolution
of a low-mass star cluster.  So, we are able to model the Galactic
tidal field as a smooth and analytic background potential.  For the
Galactic potential, we use one of two standard models.  The first
(hereafter ML from (M)iamoto-Nagai + (L)ogarithmic halo) is a
superposition of three components.  The halo is represented by a
spherical logarithmic potential of the form 
\begin{eqnarray}
  \label{eq:halopot}
  \Phi_{\rm halo}(r) & = & {1\over 2} v_{0}^{2} \ln \left( 1 +
  \frac{r^2}{d^2} \right),
\end{eqnarray}
with $v_{0}=256$~km\,s$^{-1}$ and $d=12$~kpc (and $r$ is the spherical
radius).  The Galactic disc is modelled by a Miyamoto-Nagai
potential:
\begin{eqnarray}
  \label{eq:discpot}
  \Phi_{\rm disc}(R,z) & = & - \frac{G M_{\rm d}} { \sqrt{R^{2} + \left(
        b + \sqrt{z^{2}+c^{2}} \right)^{2}}},
\end{eqnarray}
with $M_{\rm d} = 10^{11}$~M$_{\odot}$, $b = 6.5$~kpc and $c =
0.26$~kpc (where $R$ and $z$ are cylindrical coordinates).  The bulge
is represented by a Hernquist potential
\begin{eqnarray}
  \label{eq:bulgepot}
  \Phi_{\rm bulge}(r) & = & - \frac{G M_{\rm b}} {r+a},
\end{eqnarray}
using $M_{\rm b} = 3.4 \times 10^{10}$~M$_{\odot}$ and $a=0.7$~kpc. 

For comparison, we also use the Galactic potential suggested by
\citet{deh98} and hereafter denoted by DB.  It consists of three disc
components, namely the ISM, the thin and the thick disc, each of the
form
\begin{eqnarray}
  \label{eq:dehndisc}
  \rho_{\rm disc}(R,z) & = & \frac{\Sigma_{\rm d}}{2z_{\rm d}} \exp
  \left( - \frac{R_{\rm m}}{R} - \frac{R}{R_{\rm d}} -
      \frac{|z|}{z_{\rm d}} \right).
\end{eqnarray}
With $R_{\rm m} = 0$, Eq.~(\ref{eq:dehndisc}) describes a standard
double exponential disc with scale-length $R_{\rm d}$, scale-height
$z_{\rm d}$ and central surface-density $\Sigma_{\rm d}$.  For the
stellar discs, $R_{\rm m}$ is set to be zero, while for the ISM-disc,
we allow for a central depression by setting $R_{\rm m}=4$~kpc
\citep{deh98}.  In addition to the the disc potential, we use the
analytic potential corresponding to two spheroidal density
distributions for the halo and the bulge in the form
\begin{eqnarray}
  \label{eq:dehnspher}
  \rho (R,z) & = & \rho_{0} \left( \frac{m}{r_{0}}
  \right)^{-\gamma} \left( 1 + \frac{m}{r_{0}} \right)^{\gamma -
  \beta} \exp \left( - \frac{m^{2}} {r_{\rm t}^{2}} \right),
\end{eqnarray}
where
\begin{eqnarray}
  \label{eq:dehnspher2}
  m & = & \sqrt{R^{2} + \frac{z^{2}} {q^{2}}}.
\end{eqnarray}
We choose the parameters of our DB model according to the best-fit
model~4 in the paper of \citet{deh98}.  In Fig.~\ref{fig:potcomp}, we
compare the two potentials.  For both, the circular velocity at the
solar radius is $220$~km\,s$^{-1}$.  However, the ML model contains
more mass within a given radius than the DB model.

\subsection{Initial Model and Orbit for NGC~5466}
\label{sec:initial}

As a initial model for the star cluster, we choose a \citet{plu11}
sphere:  
\begin{eqnarray}
  \label{eq:plummer}
  \rho(r) & = & \frac{3 M_{\rm pl}} {4 \pi R_{\rm pl}^{3}} \left( 1 +
    \frac{r^{2}} {R_{\rm pl}^{2}} \right)^{-5/2},
\end{eqnarray}
with $R_{\rm pl}$ being the scale-length of the Plummer sphere, which
is identical to the half-light radius, and $M_{\rm pl}$ the total
mass.  This is a fairly good representation of a star cluster,
especially a young one.  However, due to tidal shaping and internal
evolution at later stages, a \citet{Ki66} model usually fits the
photometric data better.  The advantage of a Plummer model is that all
physical quantities are analytically accessible.

The initial Plummer model has a half-light radius of $10$~pc, an
initial mass of $7 \times 10^{4}$~M$_{\odot}$ and is represented by
$10^6$ particles.  The numerical set-up of the particles is performed
using the algorithm of \citet*{Aa74}.  We checked that our initial
model is able to survive for a Hubble time by comparing our initial
configuration with the dissolution times given in \citet{Ba03} (see
their fig.~3).  As the orbit of NGC~5466 is most of the time located
far out in the halo, it is well represented by the uppermost lines in
\citet{Ba03}, giving us a dissolution time of a few Hubble times.

\begin{figure}
\begin{center}
\includegraphics[height=8cm]{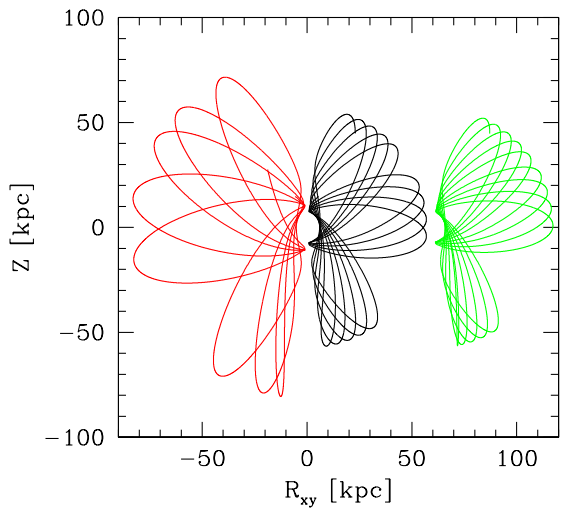}
\end{center}
  \caption{Comparison between the two orbits using the observationally
    determined proper motion. Left side (red): flipped orbit in the DB
    potential. Right side (black): orbit in the ML potential.  The
    orbit with the revised proper motions in the ML potential is shown
    in green with an off-set of $+60$~kpc.}
  \label{fig:orbit}
\end{figure}

To determine the orbit of NGC~5466, we use the positions and proper
motion from the literature \citep{Ha96,Di99}, namely:
\begin{eqnarray}
  \label{eq:posvel}
  \alpha & = & 14^{\rm h} \ 05^{\rm m} \ 27^{\rm s}\!\!.3  \ = \
  211.\!\!^{o}36 \nonumber \\ 
  \delta & = & +28^{o} \ 32' \ 04'' \ = \ 28.\!\!^{o}53 \nonumber\\
  D_{\odot} & = & 15.9 \ {\rm kpc} \nonumber \\
  \mu_{\alpha} \cos \delta & = & -4.65 \pm 0.82 \ {\rm mas}\,{\rm
    yr}^{-1} \nonumber \\ 
  \mu_{\delta} & = & 0.80 \pm 0.82 \ {\rm mas}\,{\rm yr}^{-1}
  \nonumber \\
  v_{\rm rad} & = & 119.7 \ {\rm km}\,{\rm s}^{-1} \nonumber.
\end{eqnarray}
Its Galactocentric distance is $R_{\rm GC} = 16.2$~kpc.  We transform
the positions and velocities into a Galactocentric Cartesian
coordinate system and integrate a test particle back in time for
$10$~Gyr.  This endpoint of the backward-integration is then the
starting position of our initial model.  Even though the two
potentials are quite similar in the innermost parts, the orbits differ
in terms of perigalacticon, apogalacticon and number of disc
crossings.  In the DB potential, the Galaxy is less massive in the
outer parts, so the cluster can reach an apogalacticon of $84$~kpc,
while in the ML case, it only reaches $57$~kpc.  The perigalactica are
$5$ and $6$~kpc, respectively.  In Fig.~\ref{fig:orbit}, the shape of
the orbits in the $R$,$z$-plane is plotted (for the DB model, we
flipped the radial coordinate onto the negative side to aid
visibility). 
\begin{figure}
  \begin{center}
    \includegraphics[height=7.cm]{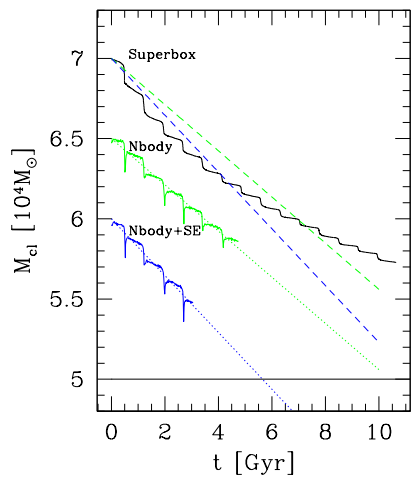}
  \end{center}
  \caption{Bound mass versus time.  Results of the NBODY~4 simulations
    in comparison with the {\it Superbox} particle-mesh simulations.
    The uppermost curve is the bound mass in the {\it Superbox}
    simulation.  The results of the NBODY~4 simulations have been
    shifted downwards by $0.5$ and $1.0$ times $10^{4}$~M$_{\odot}$,
    respectively for clarity.  The middle curve (green) shows the
    result of the NBODY~4 simulation neglecting stellar evolution and
    lowest curve (blue) shows the result of the NBODY~4 run with
    stellar evolution.  Shown are also the linear fitting lines to the
    NBODY~4 results (dashed lines).}
  \label{fig:nbody}
\end{figure}
\begin{figure}
  \centering
  \includegraphics[height=7.cm]{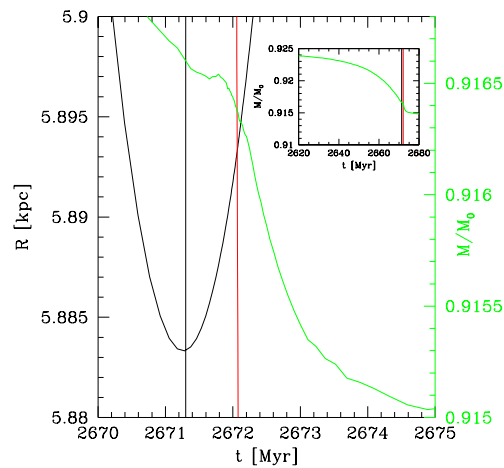}
  \caption{Mass-loss at one perigalacticon and disc-passage of the
    particle-mesh simulation in a more detailed time resolution.  The
    (black) parabolic curve shows the distance to the centre of the MW
    with the first (black) vertical line showing the time of
    perigalacticon.  The second (red) vertical line shows the time of
    the actual disc crossing ($z = 0$).  The (green) curve together
    with the right ordinate shows the evolution of the bound mass.  It
    is visible in this figure that the mass-loss ceases after
    perigalacticon and turns into a steep mass-loss at the disc
    crossing.  But as shown in the in-set (vertical lines are the
    same as in the main figure) the total mass-loss at this
    combination of perigalacticon and disc passage is $6$ times larger
    than the steep mass-loss caused by the actual disc passage.}
  \label{fig:perigal}
\end{figure}

\section{Justification of Particle-Mesh Simulations}
\label{sec:sbox}

\begin{figure*}
\begin{center}
\includegraphics[width=4.3cm]{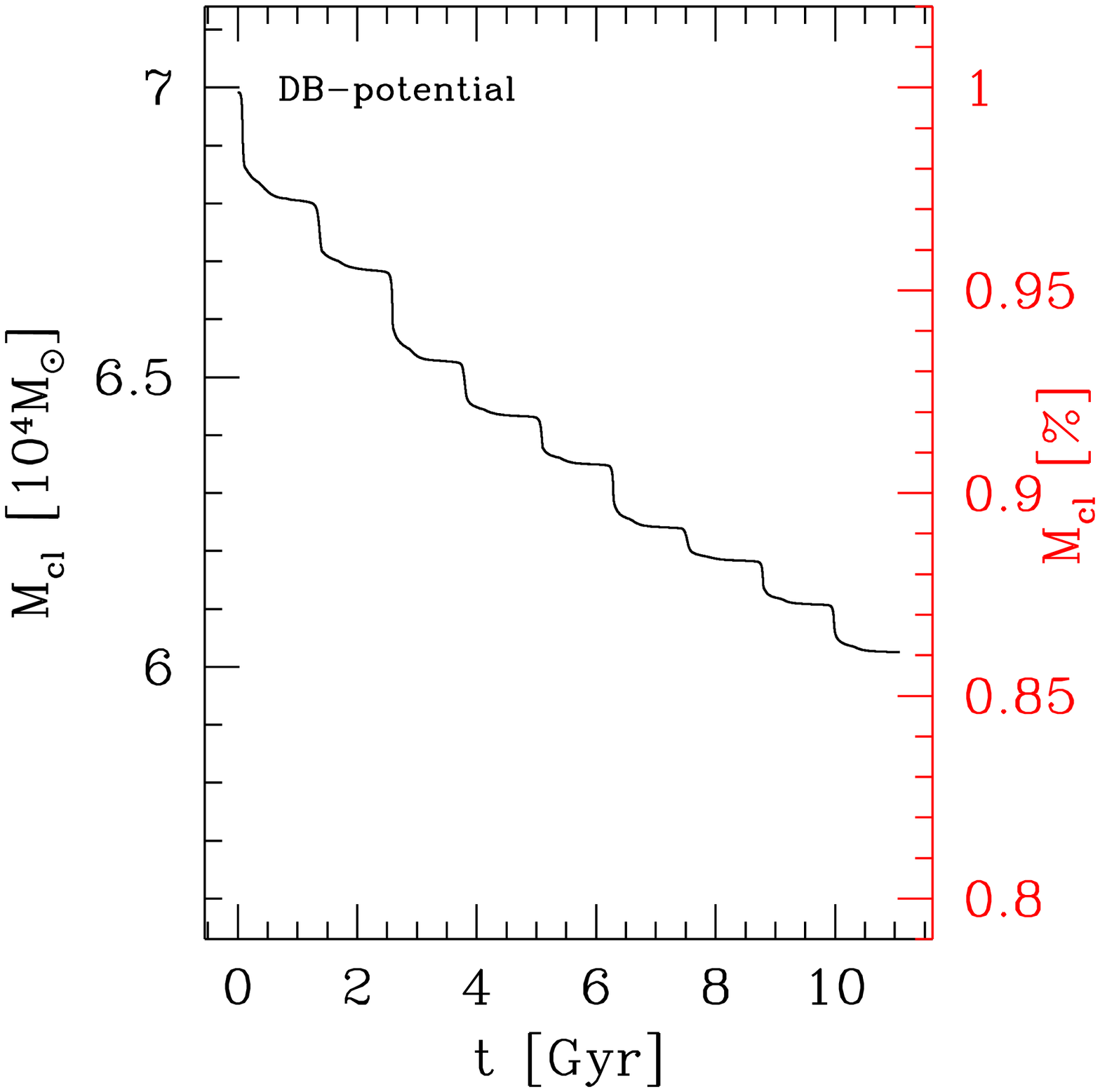}
\includegraphics[width=4.3cm]{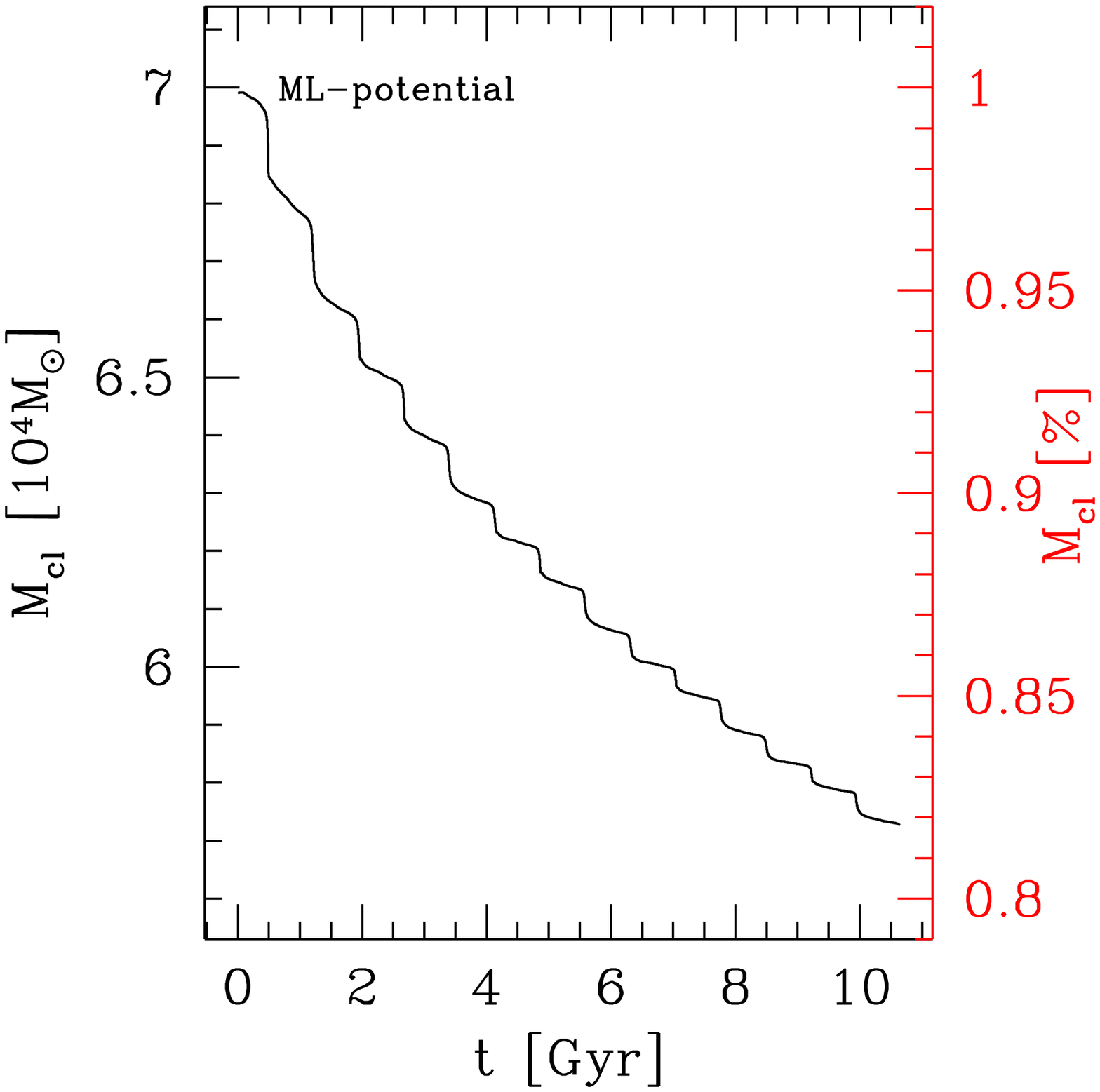}
\includegraphics[width=4.3cm]{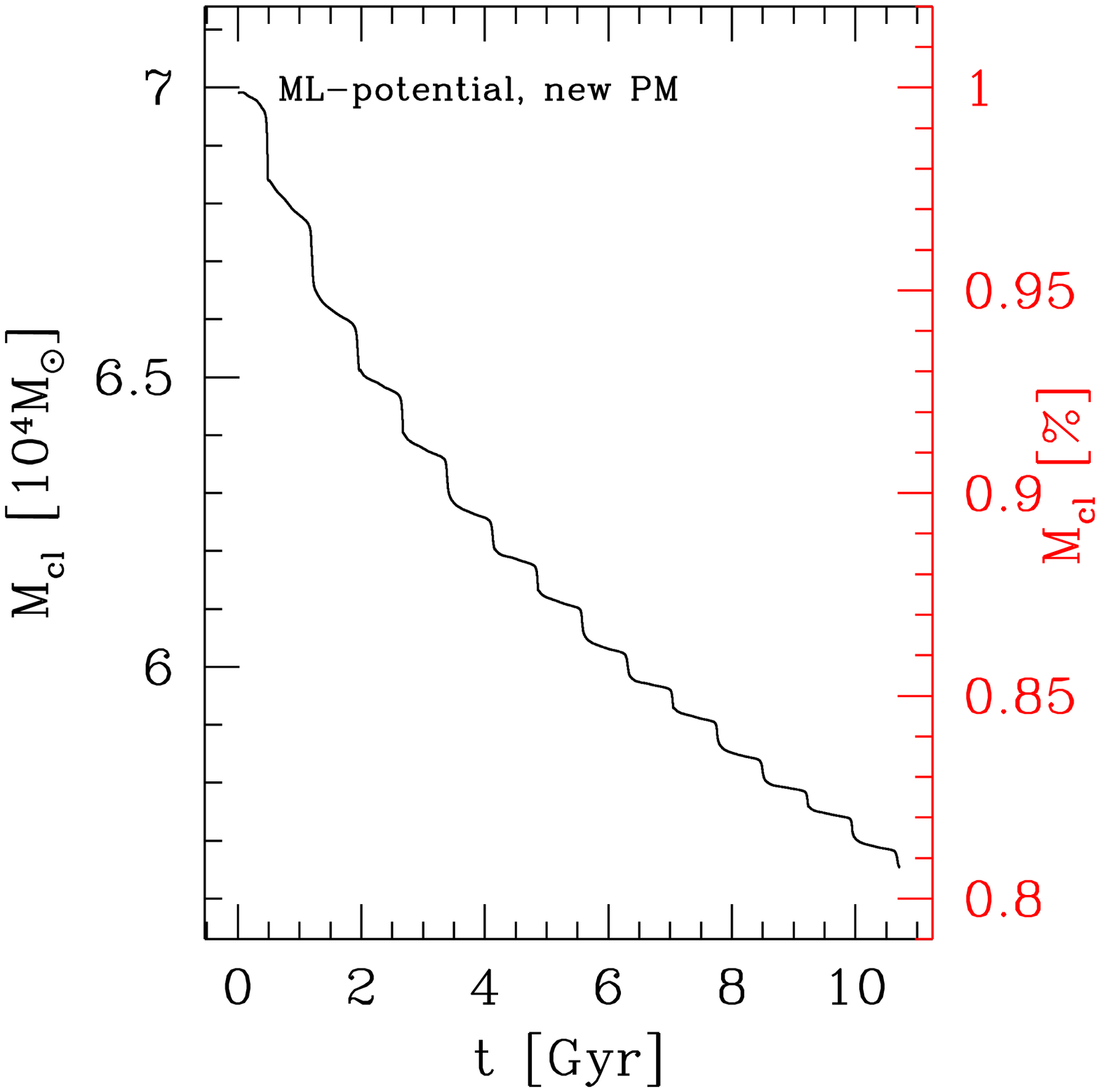}
\includegraphics[width=4.3cm]{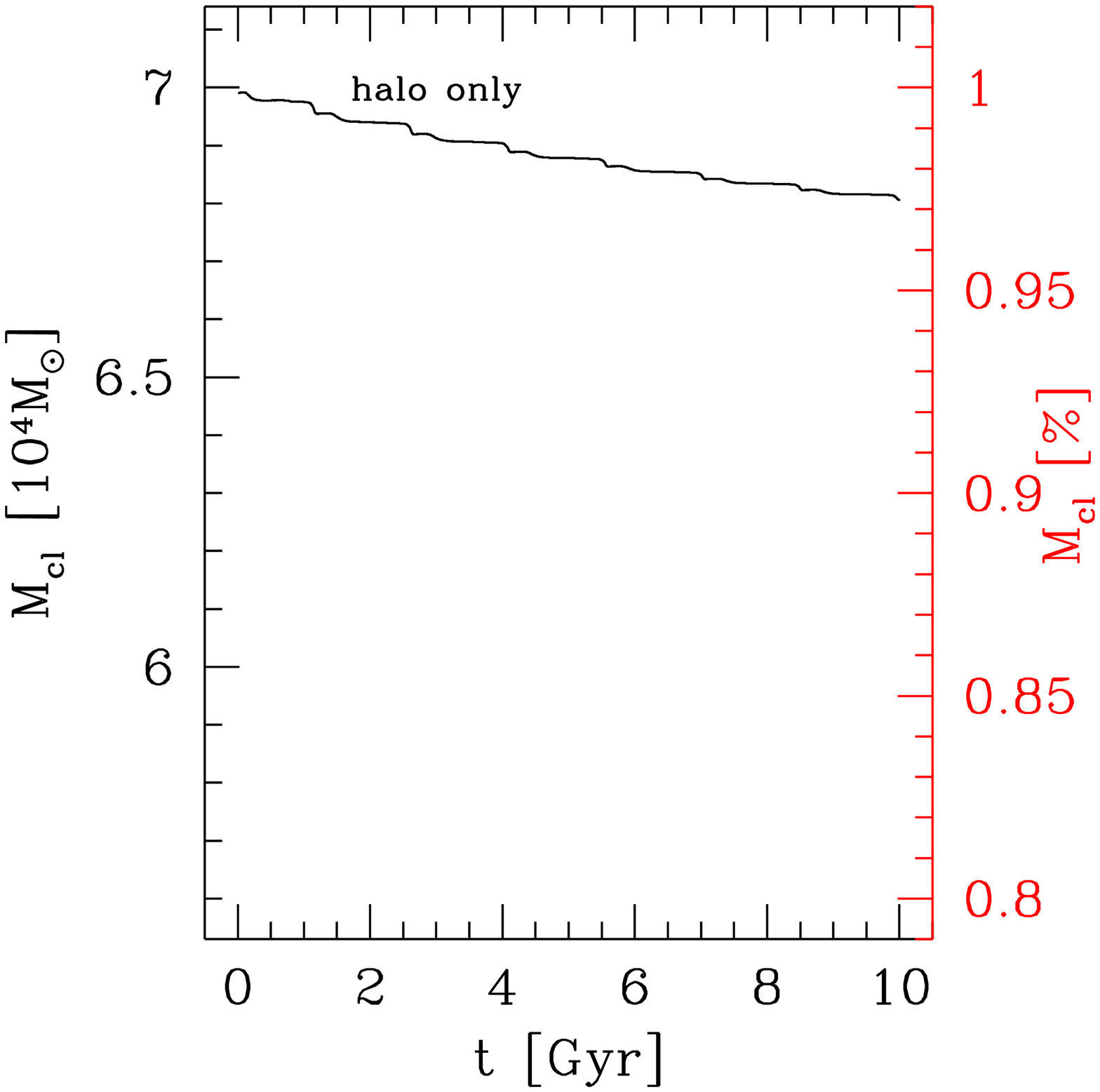}
\end{center}
  \caption{Mass-loss of NGC~5466 in the DB potential (left panel), in
    the ML potential (middle left panel), in the ML potential with the
    revised proper motions (middle right panel) and in the halo only
    potential (right panel).}
  \label{fig:mass}
\end{figure*}
\begin{figure*}
\begin{center}
\includegraphics[width=8.6cm]{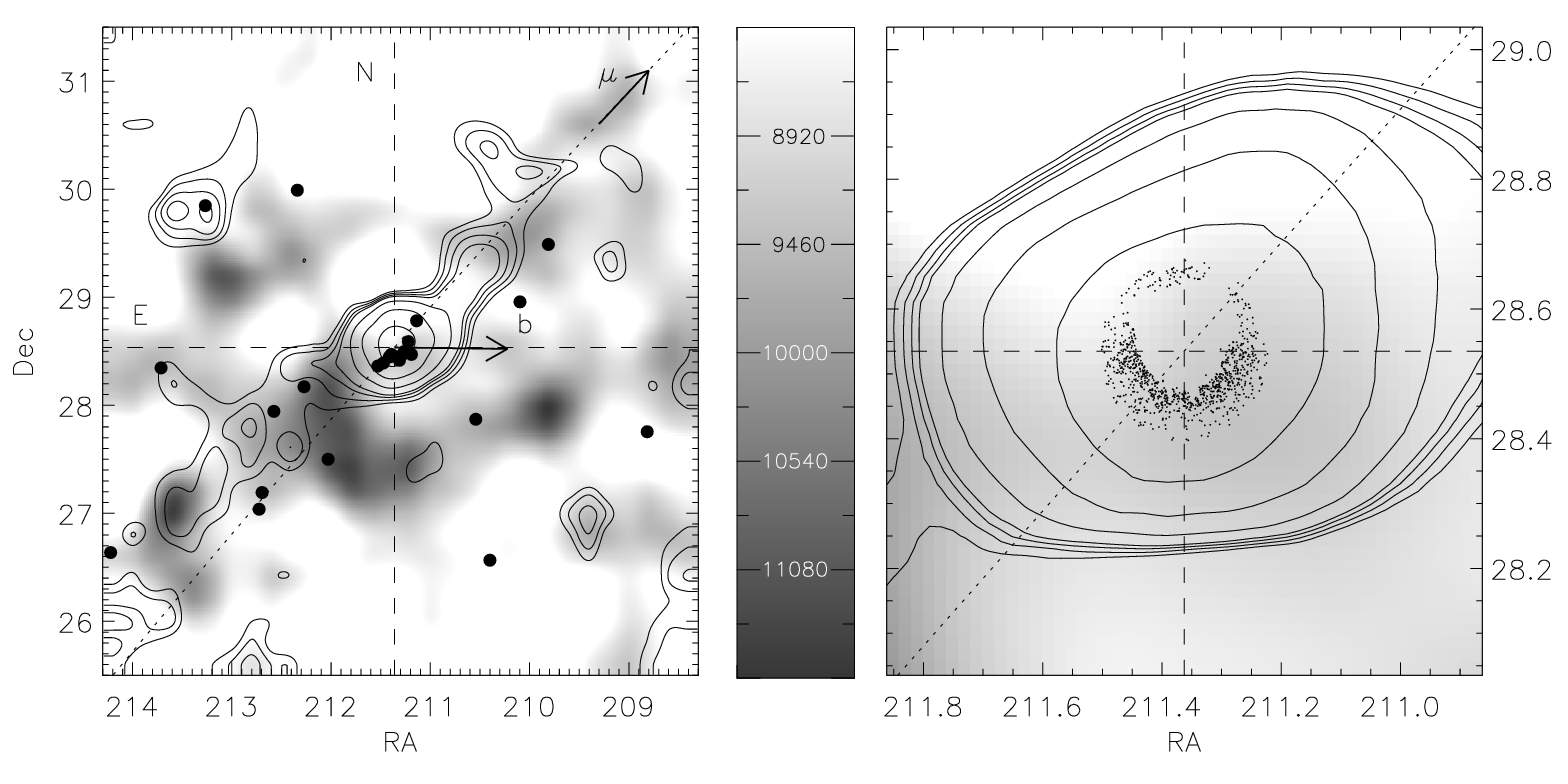}
\includegraphics[width=4.3cm]{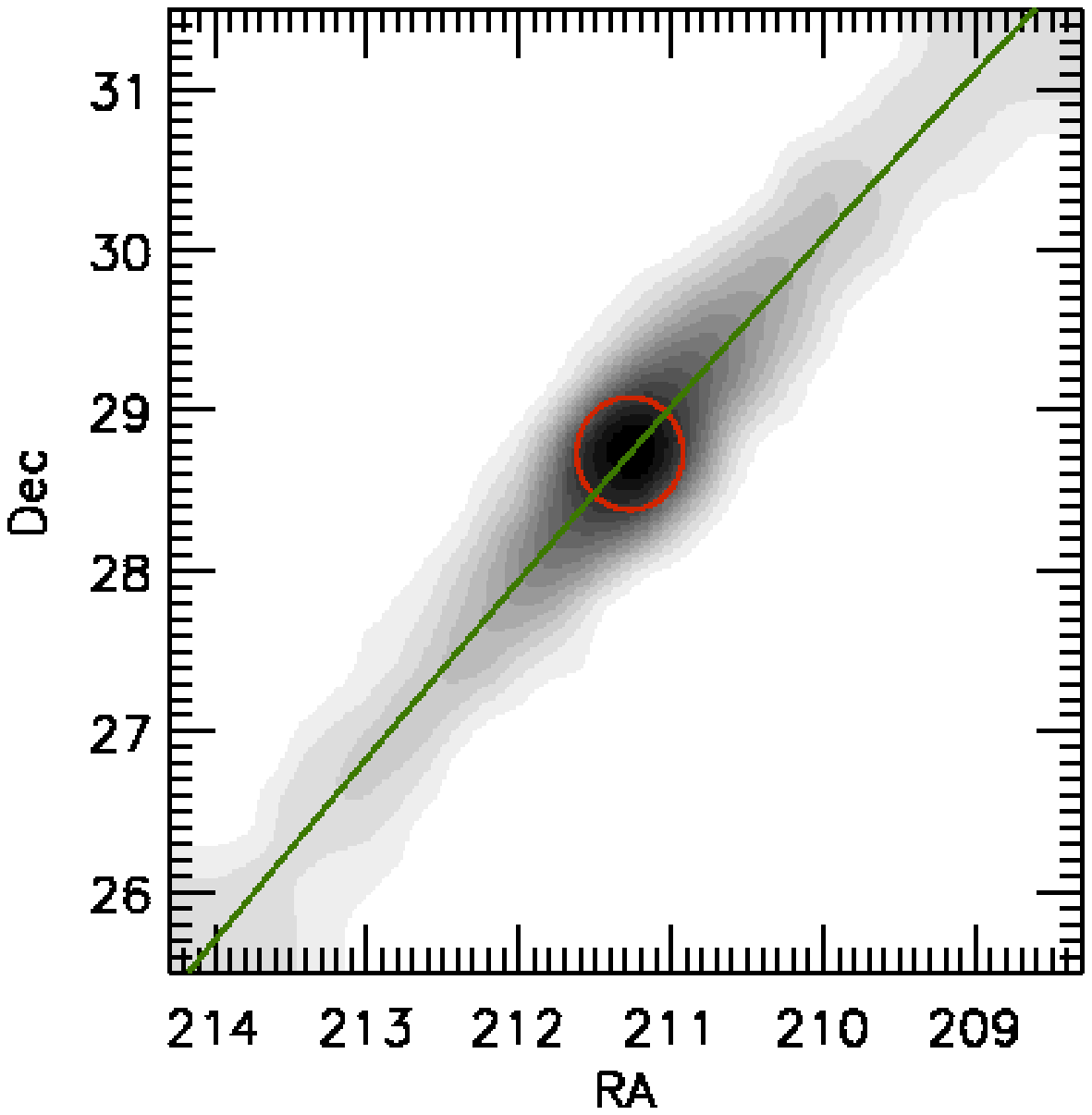}
\includegraphics[width=4.3cm]{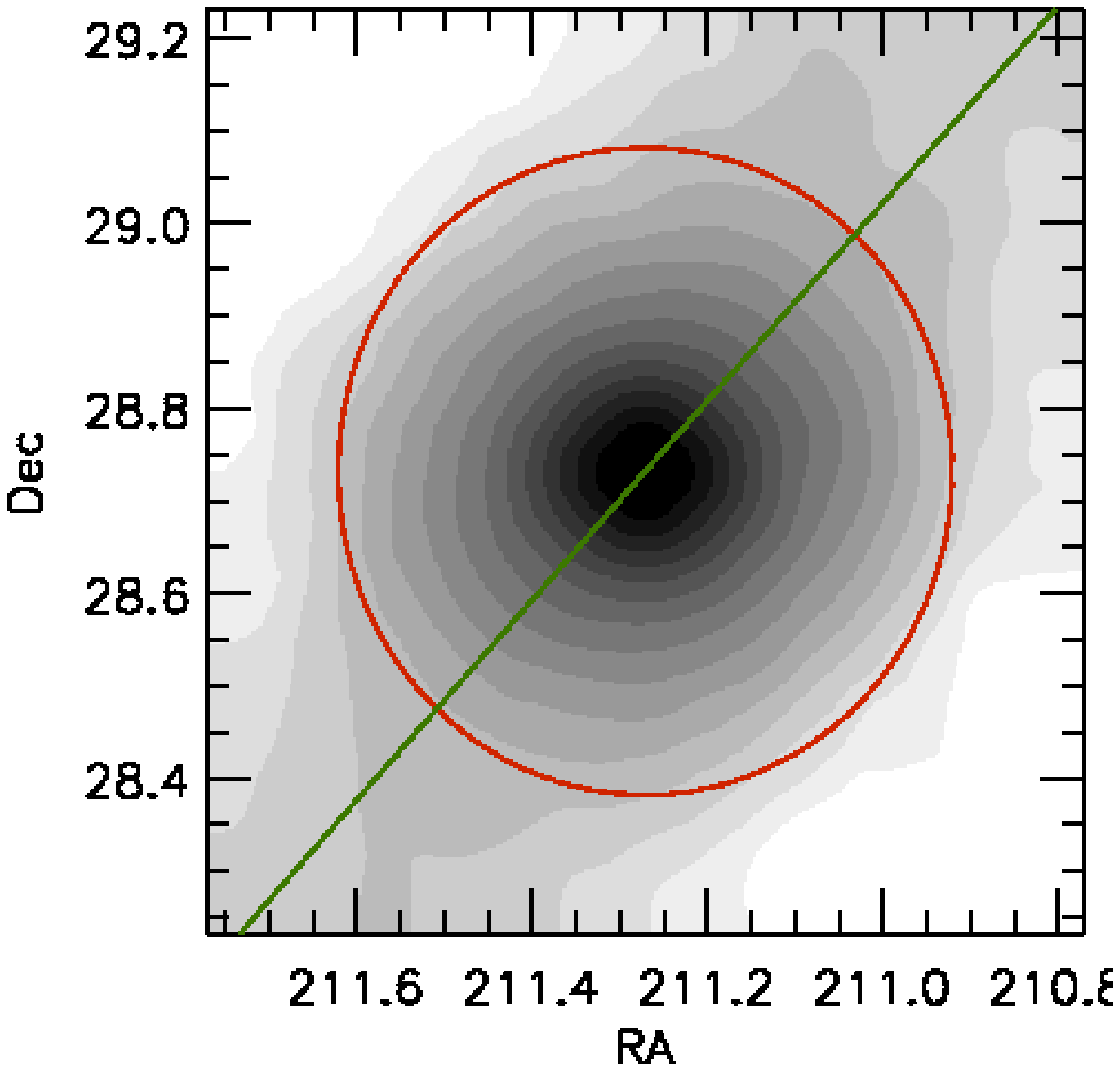}
\includegraphics[width=4.3cm]{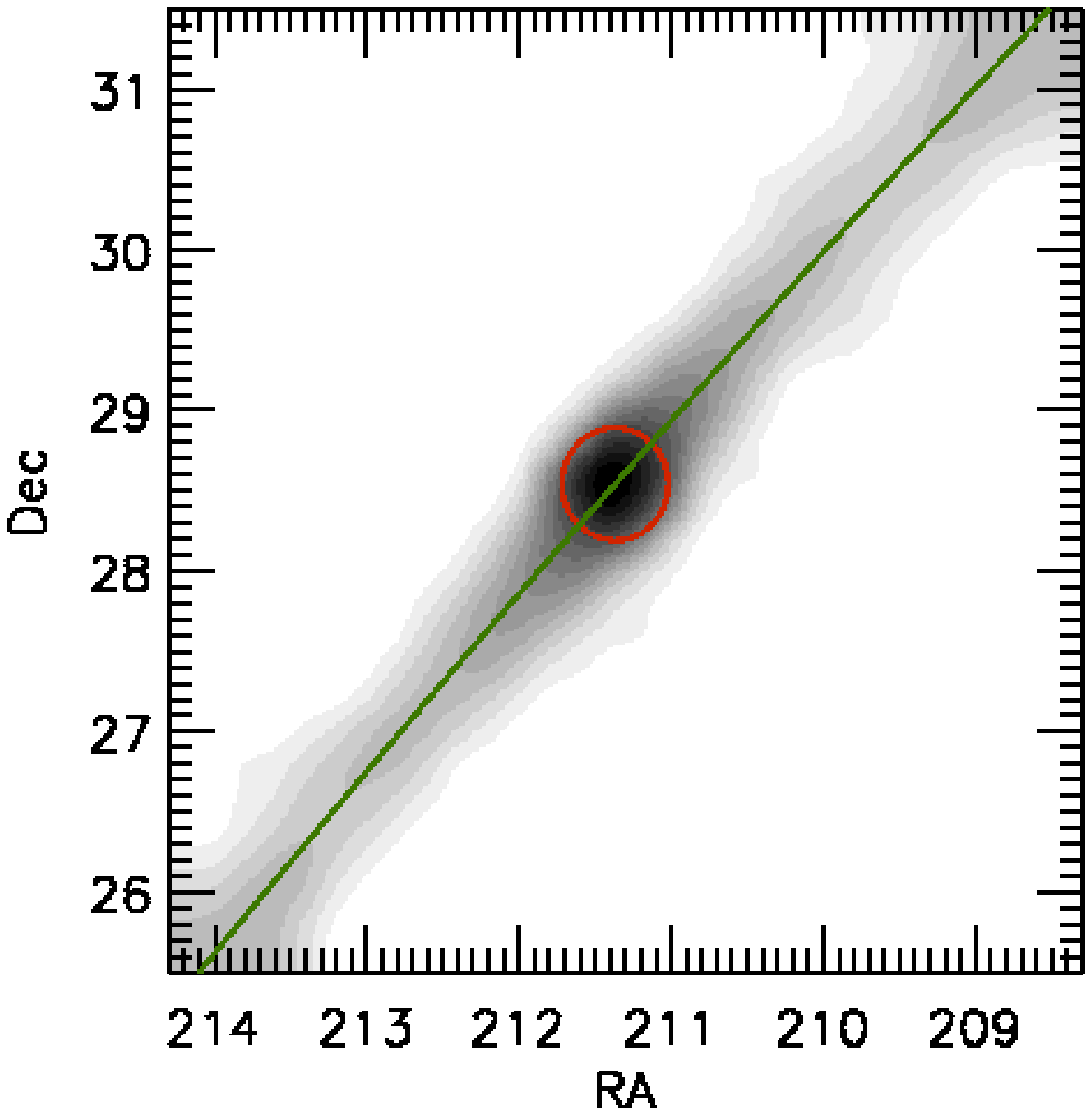}
\includegraphics[width=4.3cm]{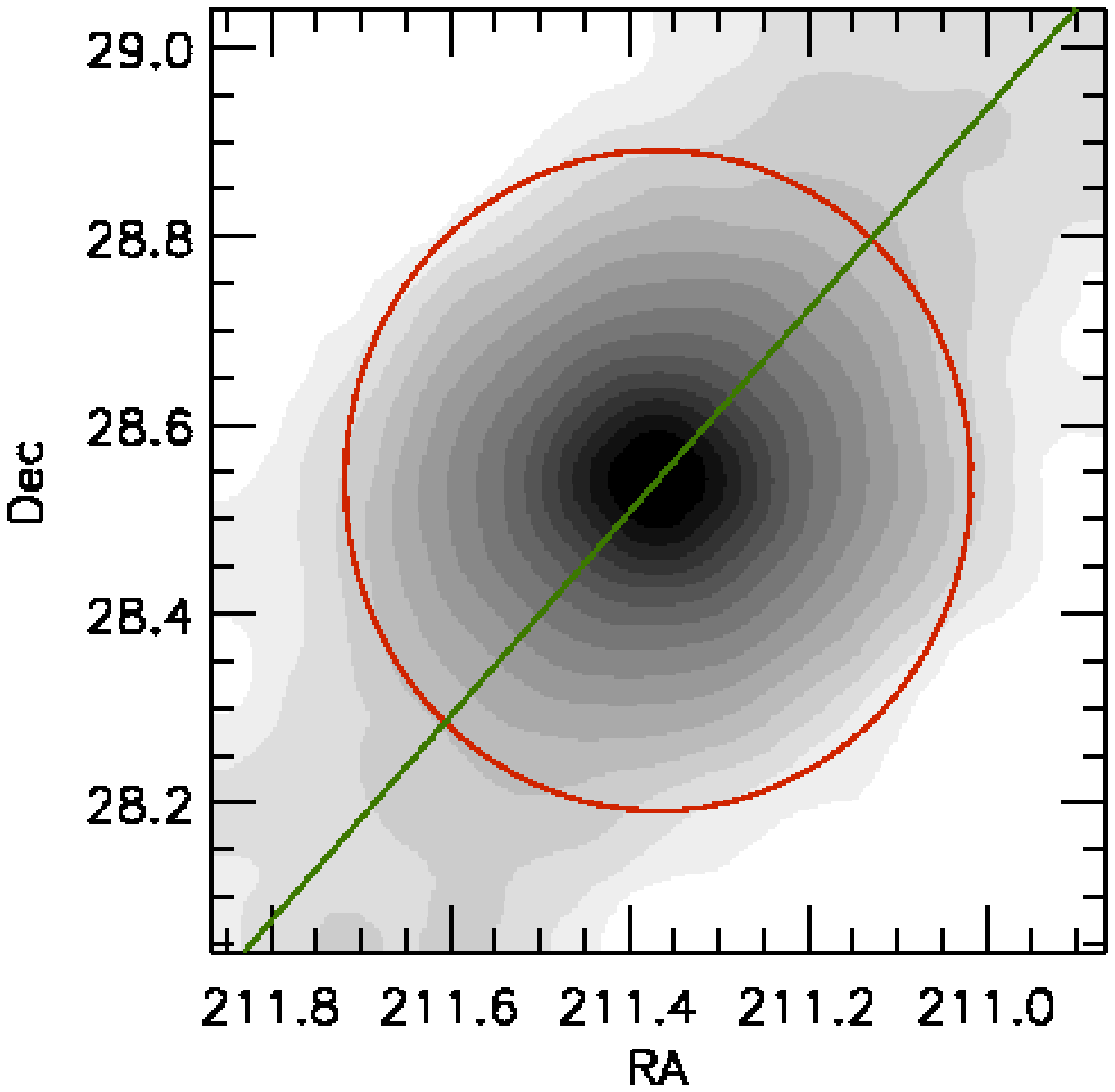}
\includegraphics[width=4.3cm]{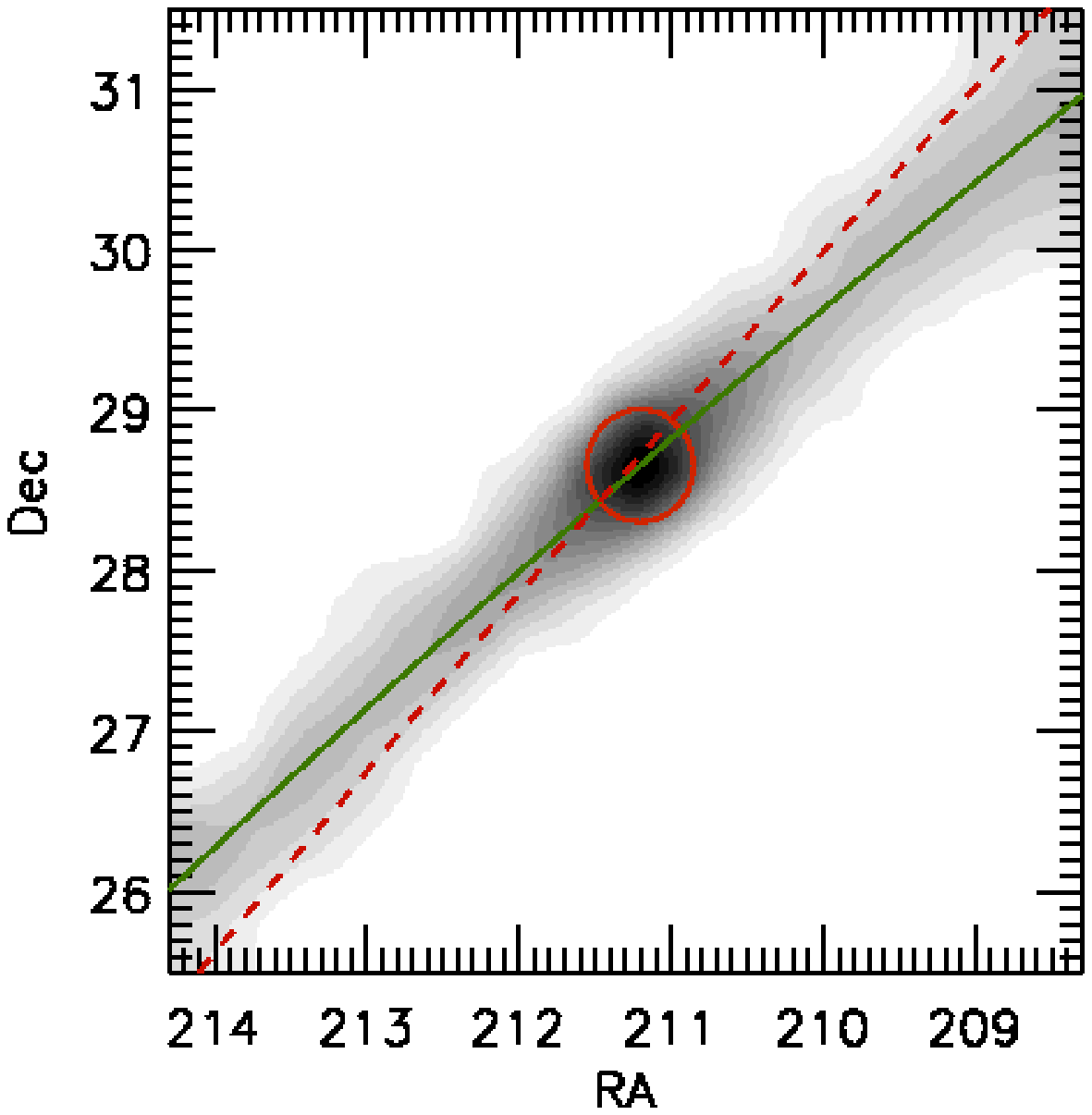}
\includegraphics[width=4.3cm]{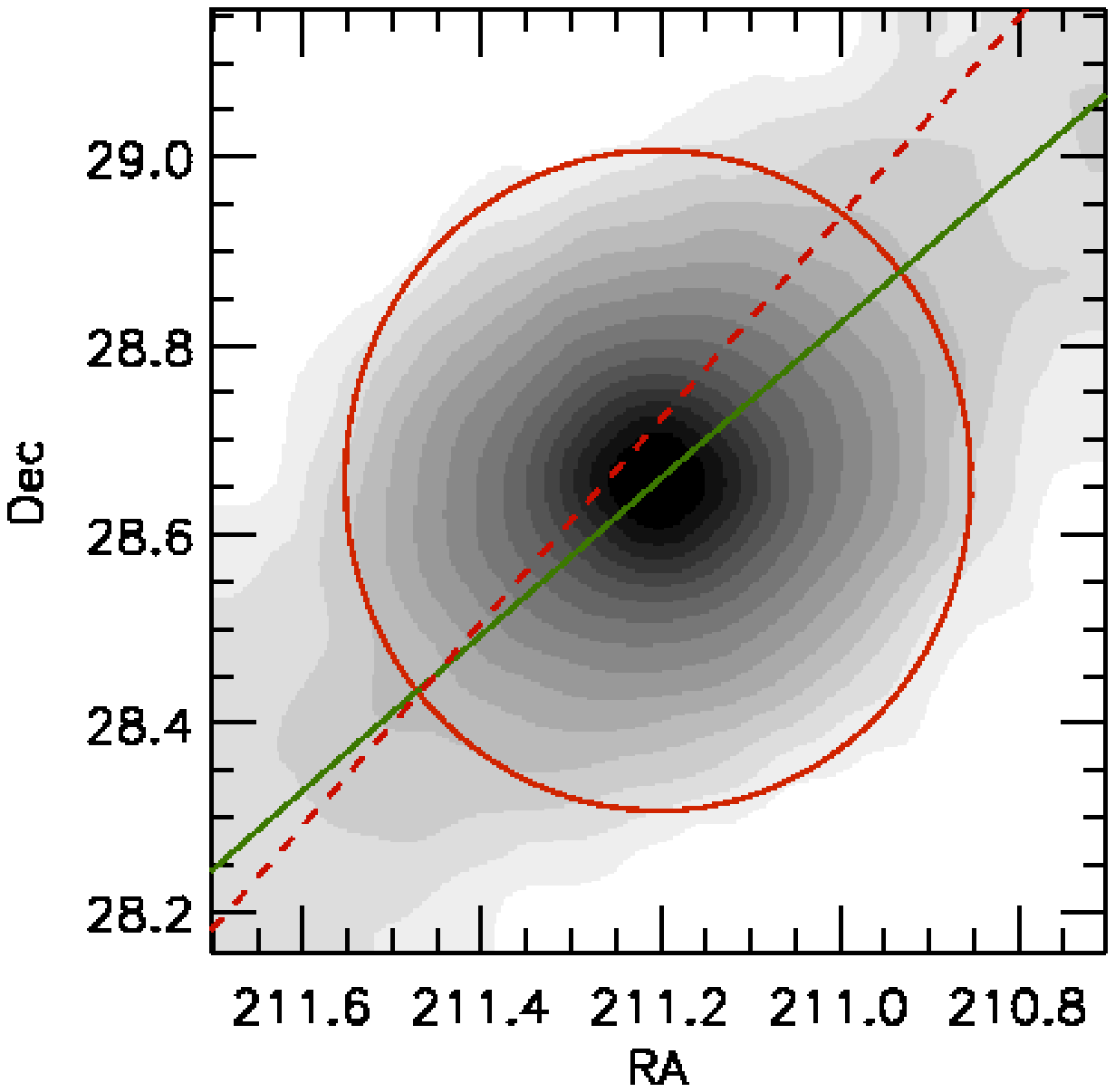}
\end{center}
  \caption{Contour plots of the tails (the model contours have
    logarithmic spacing).  The solid green line shows the actual
    orbit; the red circle the size of the actual tidal radius.  Top
    left: Observations using SDSS by \citet{Be06a}.  Top right:
    Simulation using the DB potential.  Lower left: ML potential.
    Lower right: ML potential combined with the revised proper motion.
    The tidal tails in the ML potential are more prominent than the
    those in the DB potential due to the higher mass-loss.  In all
    models, tails and orbit (solid green line) are almost aligned.  In
    both the top right and bottom left cases, the very inner tails
    are closer to the Galactic Centre in the leading arm and away from
    the Galactic Centre in the trailing arm.  This is the other way
    round in the observations.  In the lower right panel, the revised
    choice of proper motions in Eq.~(\ref{eq:revised}) is used.  Now
    the tails are a better match to the observations (compare the
    tails close to the cluster with the dashed line, which shows the
    'old' orbital path).}
  \label{fig:cont}
\end{figure*}

 To simulate the evolution of the tails of NGC~5466, we use the
particle-mesh {\it Superbox} package~\citep{fel00}. A particle-mesh
code has the great advantage that we can use millions of particles
(which represent equal-mass phase-space elements rather than single
stars) and trace the faint tails very accurately. However, such a code
is often not suitable for simulations of globular clusters, because it
neglects the internal evolution due to two-body relaxation completely.

The reason why {\it Superbox} is nonetheless a valid method for the
modelling of NGC~5466 is understood on examining the $\beta$ parameter
\citep*{Gn99}:
\begin{eqnarray}
  \label{eq:beta}
  \beta & = & \frac{t_{\rm relax}}{t_{\rm shock}}.
\end{eqnarray}
Here, $t_{\rm relax}$ denotes the relaxation time-scale, which amounts
to $\sim 3.9$~Gyr for our initial model and to $\sim 3.4$~Gyr for the
present state of the globular cluster.  Additionally, $t_{\rm shock}$
denotes the disc shock time-scale, which is the time-scale on which
the cluster is destroyed by disc shocks.  Using the formula from
\citet{Gn99}, we have
\begin{eqnarray}
  \label{eq:tshock}
  t_{\rm shock} & = & \frac{3}{4} P_{\rm disc} \frac{v^{2}\omega_{\rm
      h}^{2}} {g_{\rm m}^{2}},
\end{eqnarray}
where $P_{\rm disc}$ is the period of the disc crossings, $v$ is the
velocity with which the object crosses the disc, $\omega_{\rm h}$
denotes the ratio of velocity dispersion to half-mass radius $r_{\rm
  h}$ of the object and finally $g_{\rm m}$ is the acceleration
perpendicular to the disc.  Using our simulation data, we derive a
disc shock time-scale of about $110$~Gyr.  This gives a $\beta = 0.03
\pm 0.01$, which holds for both Galaxy models within the errors.  The
concentration 
\begin{eqnarray}
  c & = & \log\left( \frac{r_{\rm tidal}}{r_{\rm h}} \right)
\label{eq:con}
\end{eqnarray}
of our initial model and the star cluster today is in the order of
unity.  If we now place our initial model in fig.~13 of \citet{Gn99},
we see that it falls in the regime where shocks are more important
than internal evolution, but also in the regime where the star cluster
survives for a Hubble time.  Still, the location of our model is close
to the border-line (at $c \approx 1$ it is $\beta = 0.01$) where
internal evolution becomes dominant. It is interesting to compare
NGC~5466 with the well-studied case of Pal~5, which has $c = 0.6$ and
$\beta = 10$. Pal 5 will most likely be destroyed at its next disc
crossing~\citep{De04}. By contrast, NGC~5466 has a good chance of
surviving even for the next Hubble time!

To demonstrate that the internal evolution has no major effect, we
perform two N-body simulations with $10^5$ particles in the ML
potential using NBODY4~\citep{Aa99} and compare to the {\it Superbox}
results.  In the first simulation, we use an equal mass for all
particles and neglected stellar evolution.  In the second simulation,
we adopt a mass function which is present after the initial phase of
violent mass-loss caused by the evolution of high mass stars (first
few tens of Myr).  In practice, this means another $20$~\% has to be
added to the initial mass to account for the mass-loss due to
supernovae, and stellar winds, as well as the stars which become
unbound due to this mass-loss.  For the remaining stars, stellar
evolution in NBODY~4 is switched on.  Figure~\ref{fig:nbody} shows, by
extrapolating the mass-loss in the direct N-body simulations linearly,
that the additional mass-loss due to two-body relaxation and stellar
evolution amounts at the very most to about one-third of the total
mass-loss. The linear extrapolation of the mass-loss in this mass
regime is justified by appeal to the work of \citet{Ba03}.  In other
words, disregarding the initial mass-loss when the star cluster blows
away its gaseous envelope (which depends mainly on the star formation
efficiency) and the violent stellar evolution in the first few tens of
Myr, the dominant cause of mass-loss during the long-term evolution of
NGC~5466 is the tidal field of the MW.

Having established that internal disruptive processes (e.g. two-body
relaxation) are of minor importance, we also have to prove if disc
shocks are the major external destruction process.  We therefore
performed a particle-mesh simulation with a halo-only potential.  The
mass-loss in this case amounts to $3$~\% of the initial mass only (see
Fig.~\ref{fig:mass} right panel).  This is a factor of $6$ less than
in the combined potential case.  But this is not yet a genuine proof
of the importance of disc shocks.  As one can see in
Figs.~\ref{fig:nbody} and~\ref{fig:mass} the mass-loss happens in
short time-intervals like a step function.  In Fig.~\ref{fig:perigal}
we blow up one of these short time intervals.  The first (black)
vertical line shows the time of perigalacticon while the second (red)
line shows the time when the cluster crosses the disc ($z = 0$).
While the mass-loss due to the tidal field ceases after perigalacticon
there is an additional steep mass-loss starting when the cluster
passes the centre of the disc.  But shown in the small in-set in
Fig.~\ref{fig:perigal} this mass-loss is about $1/6$ of the total
mass-loss at the combined perigalacticon and disc passage.  The
general conclusion is therefore that it is definitely the tidal field
of the disc (the perigalacticon is well outside the bulge region)
which causes the major contribution of the mass-loss, the actual disc
shock when the cluster passes through the centre of the disc may be
not that important.  This finding explains the rather large time-scale
for disc shocks ($110$~Gyr) of the previous section.

\section{Tidal Tail Results}
\label{sec:res}

\subsection{The Tail Morphology and the Proper Motion}

One of the advantages of {\it Superbox} is that it has high resolution
sub-grids, which stay focused on the simulated objects and travel with
them through the simulation area. This is important in studying the
morphology of the tenuous and diffuse tidal tails.  Within the
innermost grid, we resolve the globular cluster at a resolution of
$1.7$ pc.  The grid with medium resolution is chosen to resolve the
tidal tails close to NGC~5466 with a resolution of $16.7$ pc.

In Fig.~\ref{fig:mass} (first two panels), we plot the bound mass of
our models in two Galactic potentials (DB and ML).  In both cases, the
mass-loss is strong in the first $2$--$3$ Gyr and then tends to level
off during the later stages of the evolution.  The mass-loss is mainly
related to each disc crossing near perigalacticon and the mass stays
almost constant during the rest of the orbit.  This is the major
reason why in the DB potential the mass-loss over $10$~Gyr of
evolution is less than in the ML potential -- the star cluster has had
fewer disc crossings.  In the ML potential, the cluster loses about
$18$~\% of its initial mass, whilst the cluster in the DB potential
only suffers a mass-loss of $14$~\% (with a fluctuation of only a few
particles out of 1,000,000).  The mass-loss of the ML simulation is in
very good agreement with the results found in previous studies by
\citet{He61} and \citet{Le87}.  However, these mass-losses are only
lower limits, as there will be a smaller, but not negligible,
contribution from internal relaxation effects.

\begin{figure}
  \begin{center}
    \includegraphics[height=2.7cm]{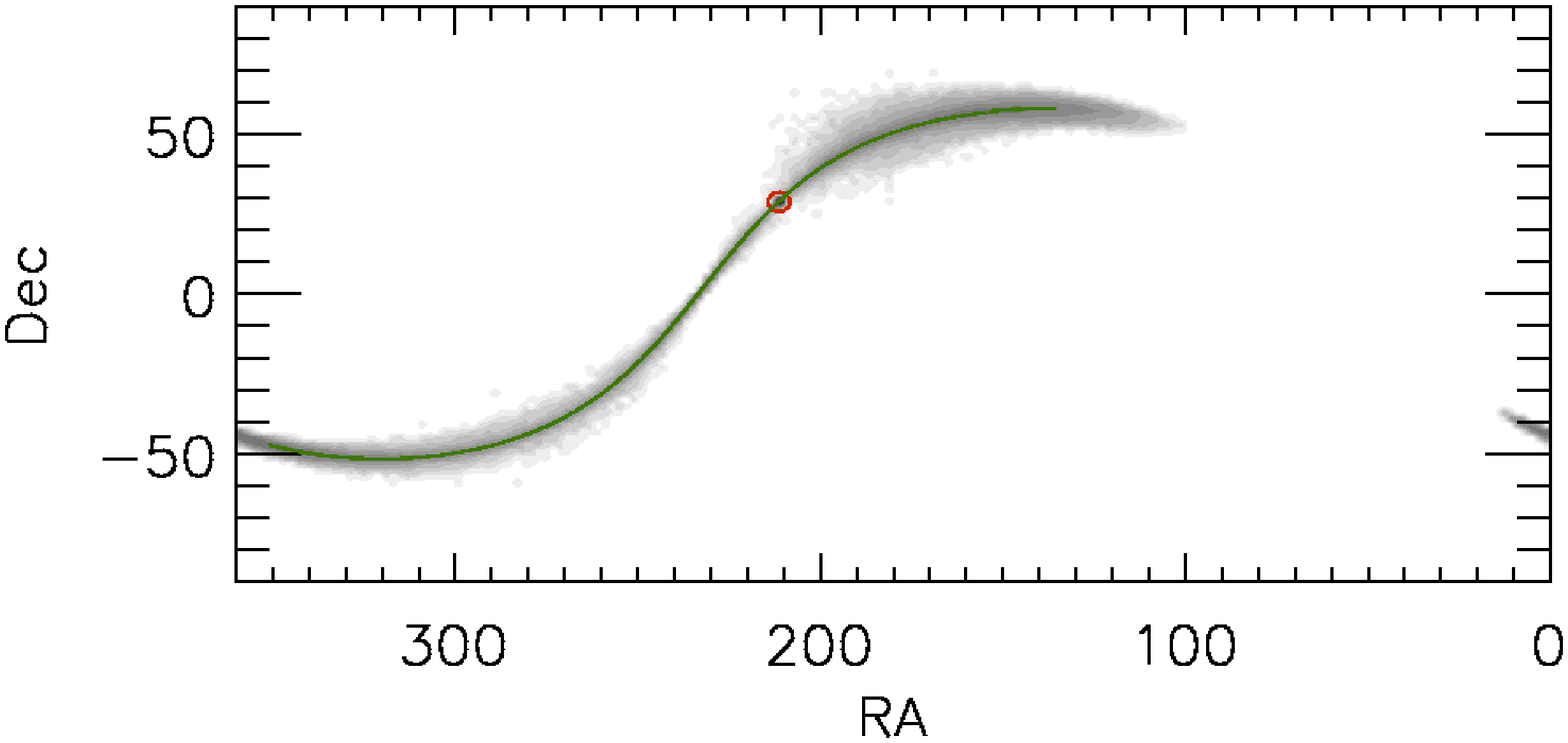}
    \includegraphics[height=2.7cm]{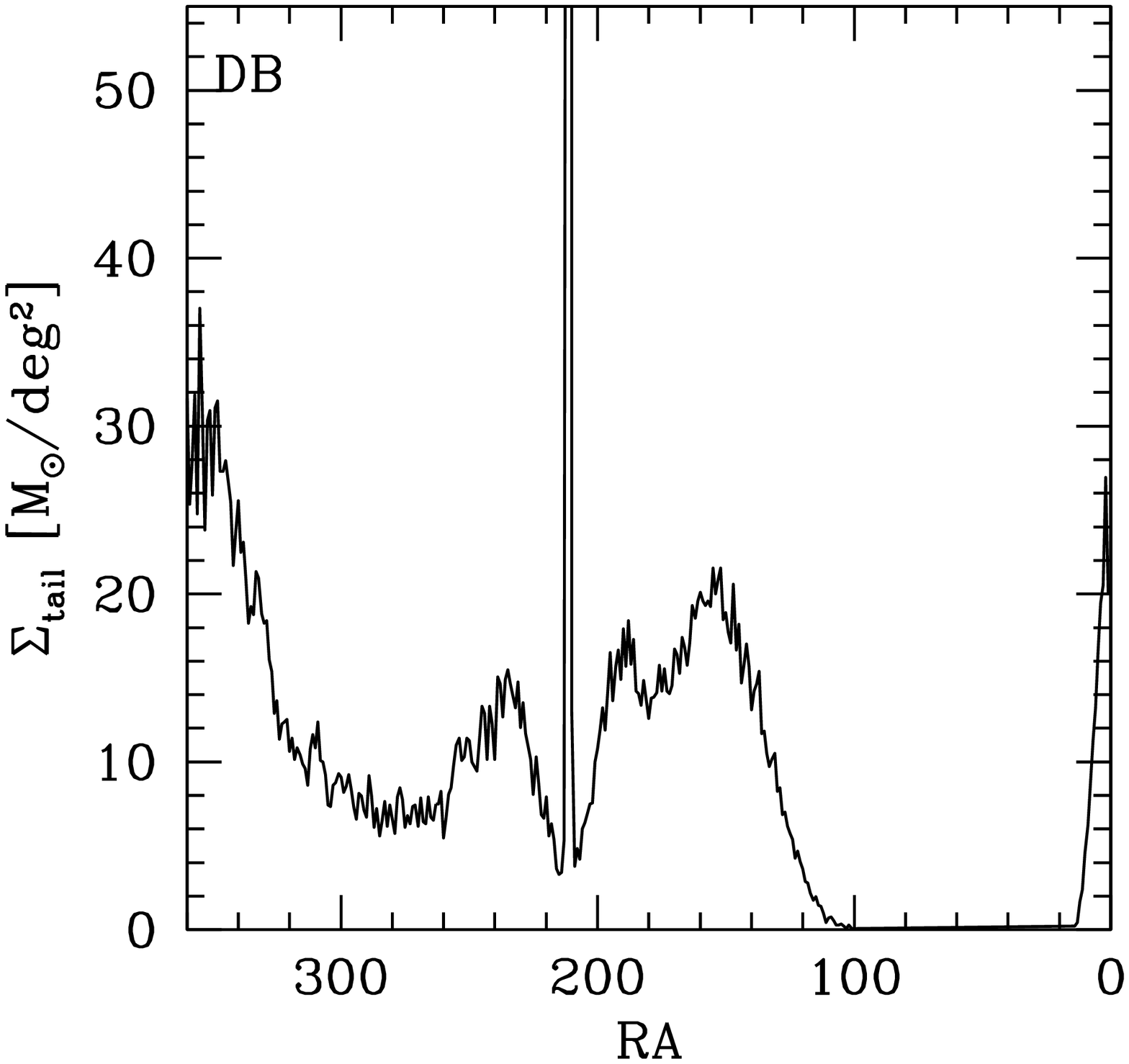}
    \includegraphics[height=2.7cm]{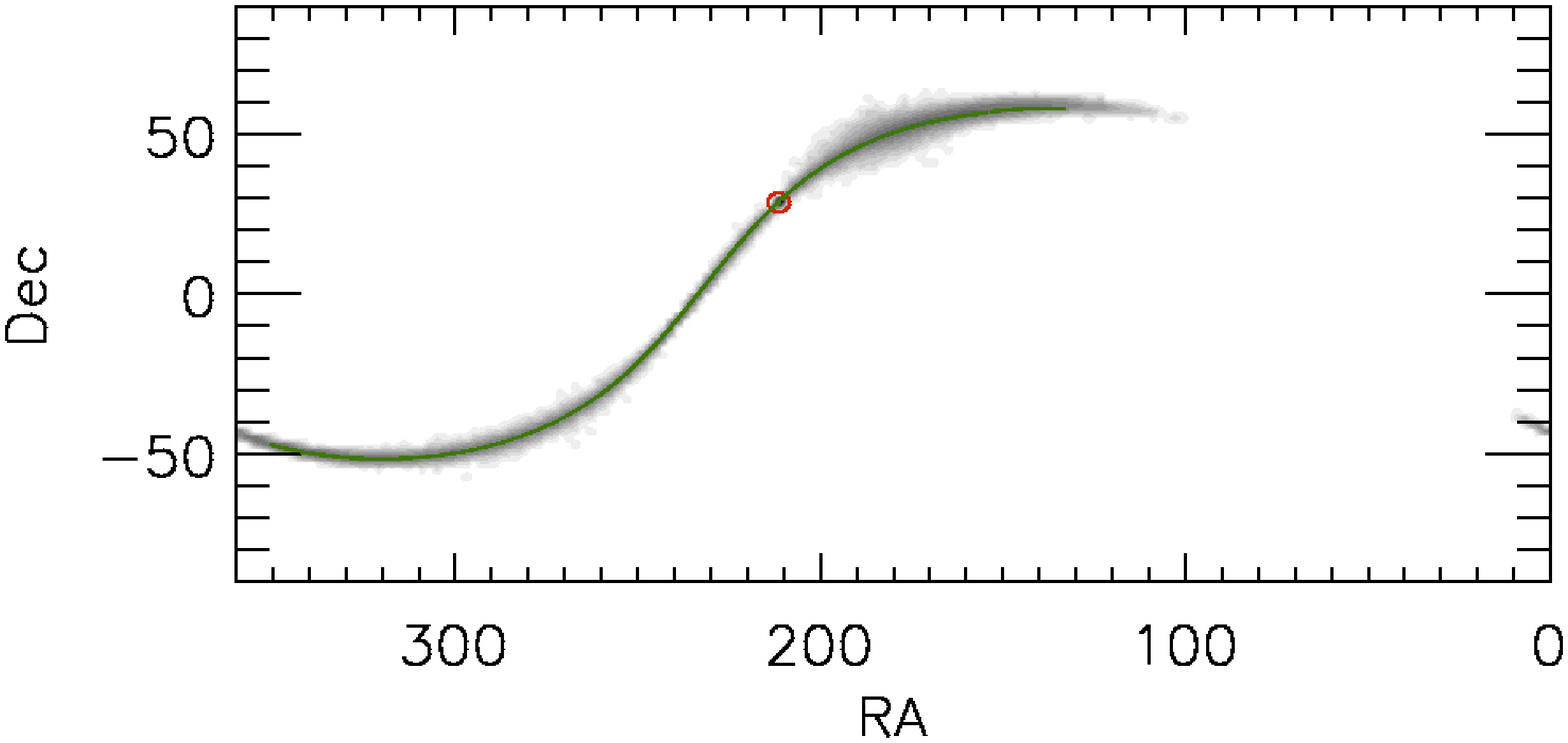}
    \includegraphics[height=2.7cm]{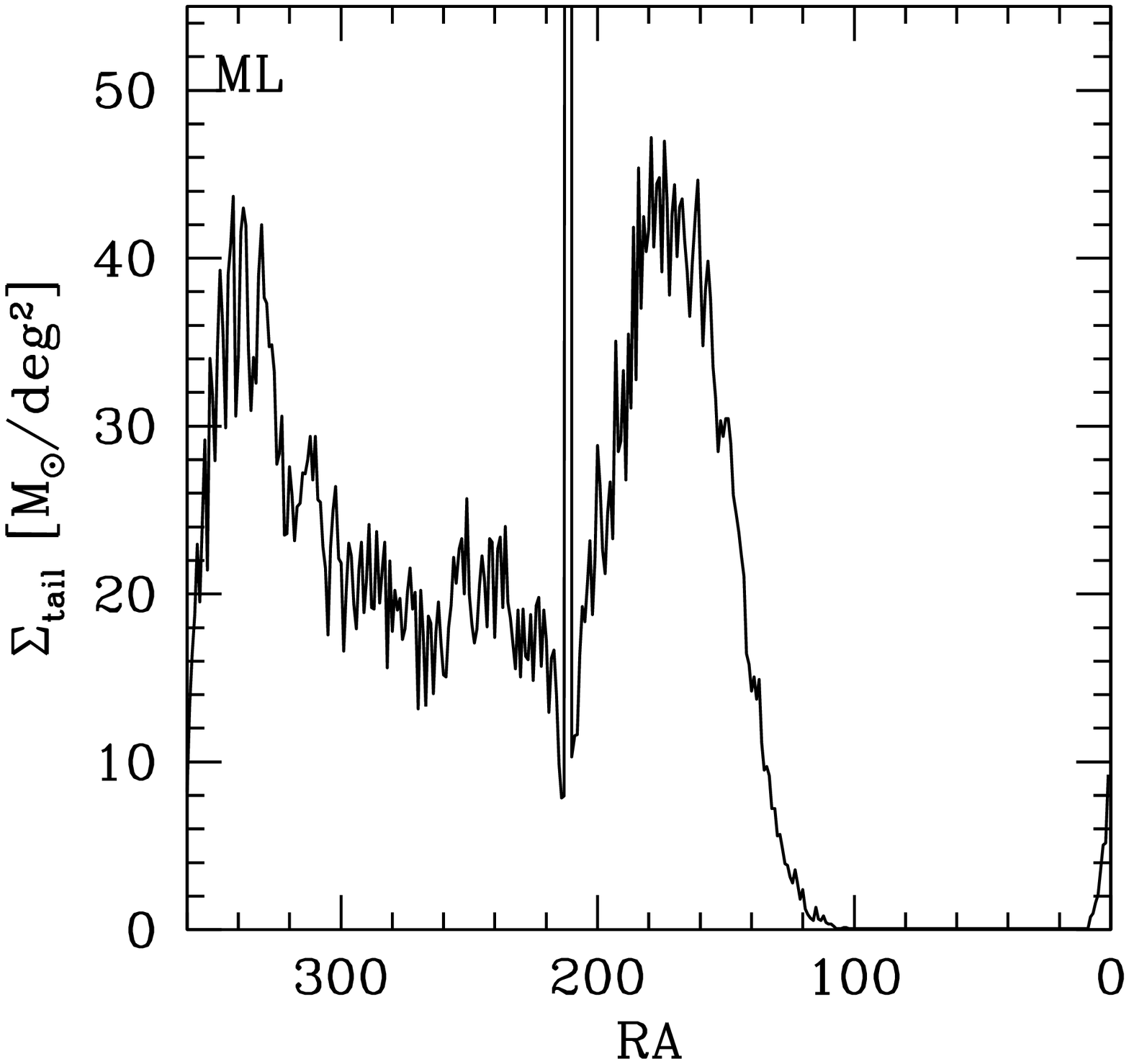}
    \includegraphics[height=2.7cm]{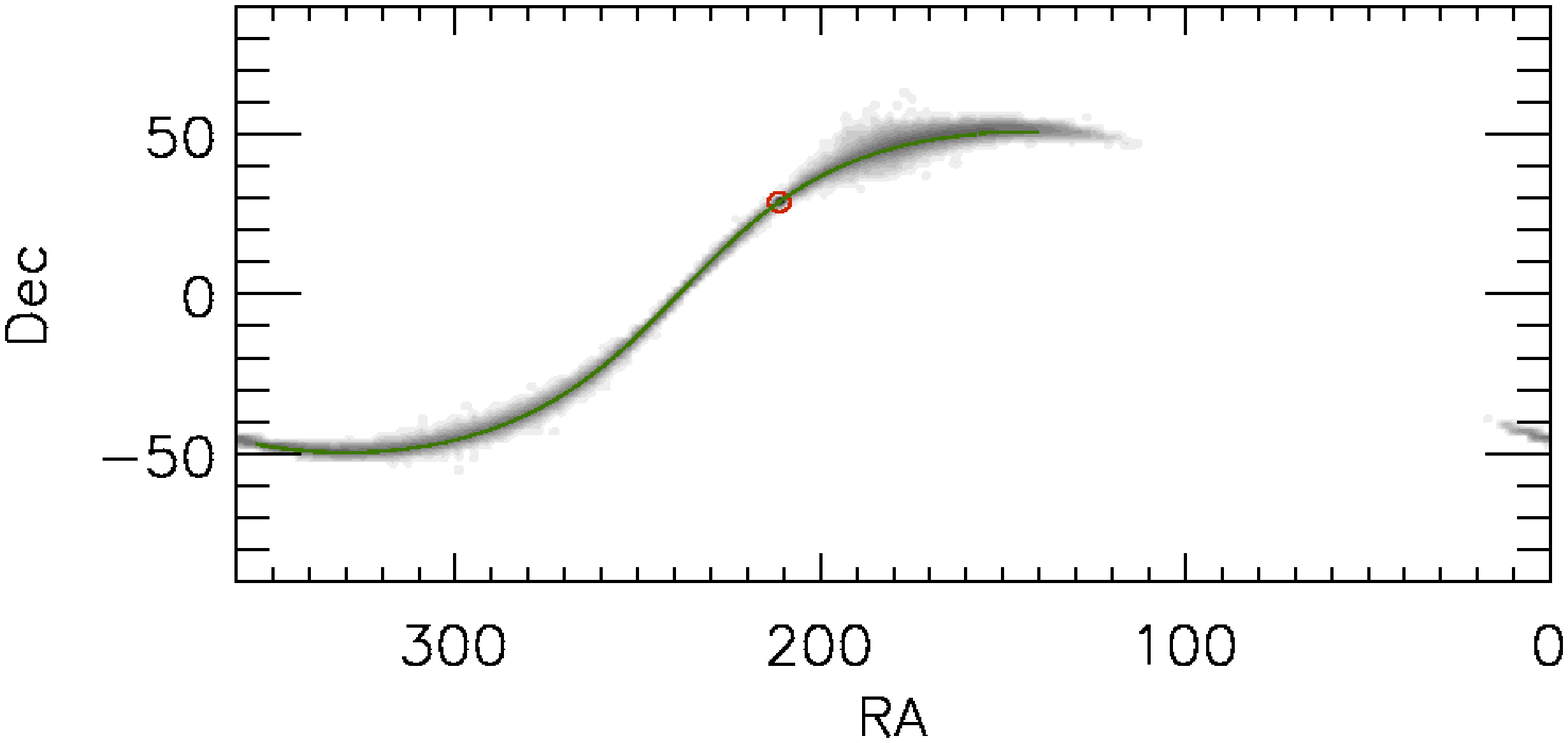}
    \includegraphics[height=2.7cm]{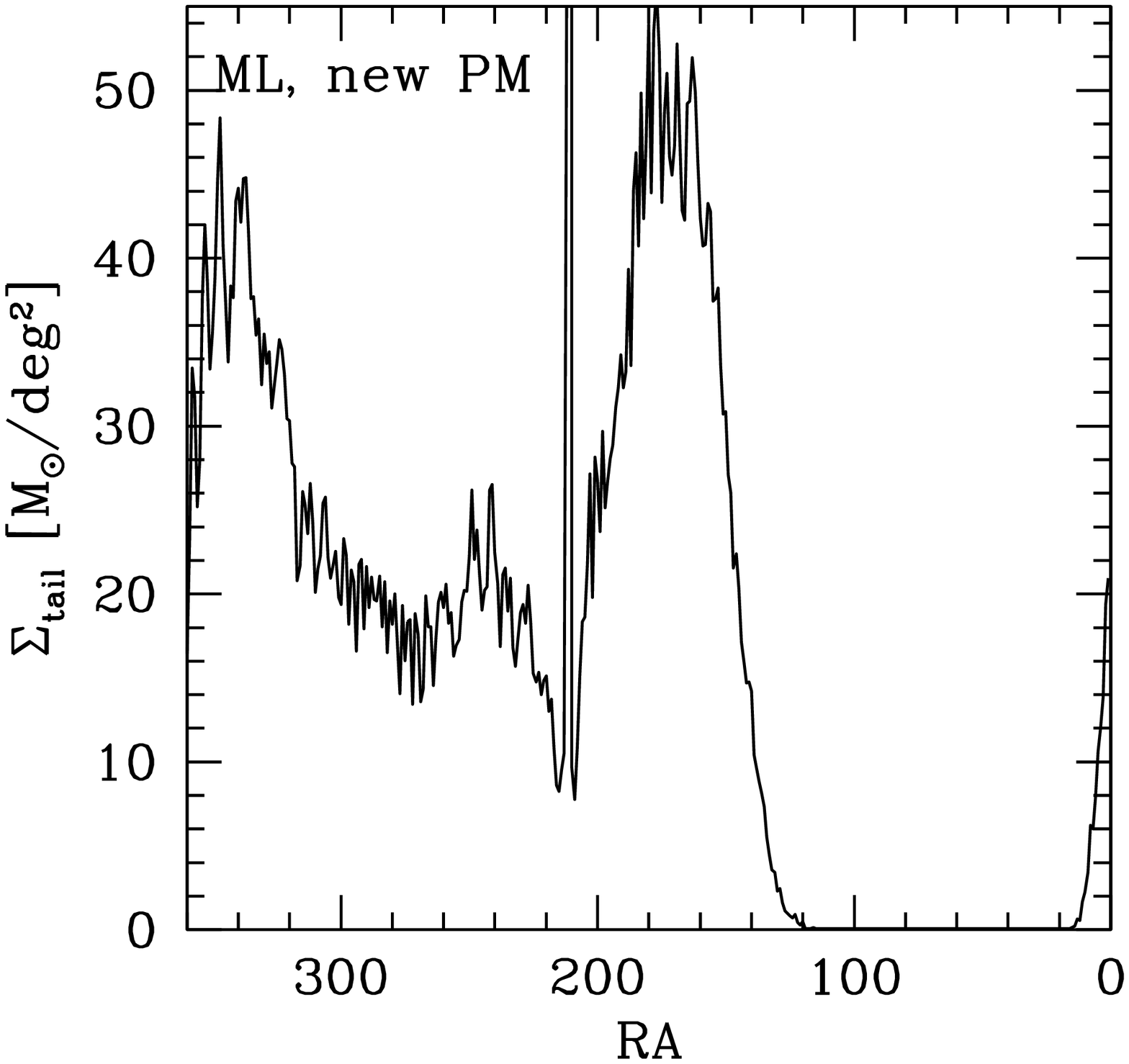}
  \end{center}
  \caption{Left panels: All-sky view of the tidal tails in our
    simulations.  The solid green line shows the orbit of NGC~5466,
    its position today is marked with a red circle.  Right panels:
    Maximum surface density along the tail.  The sky is cut into
    square degrees in right ascension and declination and for each
    degree in right ascension the maximum surface density of all
    declinations is given.  From top to bottom we show the results of
    the DB, ML and ML with revised proper motion models.}
  \label{fig:allsky}
\end{figure}

In the top left panel of Fig.~\ref{fig:cont}, we show the data on the
tails of NGC~5466 reproduced from \citet{Be06a}, who used neural
networks are used to reconstruct the probability density distribution.
The contours correspond to level curves of equal neural network output
and therefore trace the star density. The tails are clearly visible
once the extragalactic contaminants (predominantly galaxy clusters)
have been eliminated. The tails extend $\sim 4^\circ$ on the sky,
corresponding to $\sim 1$ kpc in projected length.

\begin{table*}
  \centering
  \caption{Results of the suite of simulations investigating the
    relation between orbit and density of the tidal tails.  The
    columns give the following numbers: the absolute value of the
    proper motion, the proper motion in $\alpha$ and $\delta$, the
    peri- and apogalacticon distances, the mean density of the tails,
    the maximum density in the tails, the extent of the tails, and the
    final mass of the cluster in units of the initial mass.}
  \label{tab:pm}
  \begin{tabular}{@{}ccccccccc}
    $\left| \mu \right|$ & $\mu_{\alpha} \cos \delta$ & $\mu_{\delta}$
    & $R_{\rm peri}$ & $R_{\rm apo}$ & $\Sigma_{\rm mean}$ &
    $\Sigma_{\rm max}$ & Extent & Final cluster \\ 
    mas\,yr$^{-1}$ & mas\,yr$^{-1}$ & mas\,yr$^{-1}$ & kpc & kpc &
    M$_{\odot}$\,deg$^{-2}$ & M$_{\odot}$\,deg$^{-2}$ & deg & mass \\ \hline 
    4.40 & -4.40 & 0.00 & 4.9 & 42.9 & 30.6 & 81.5 & 249 & 0.74 \\
    4.61 & -4.60 & 0.30 & 5.7 & 50.8 & 26.7 & 60.5 & 244 & 0.79 \\
    4.72 & -4.70 & 0.42 & 5.9 & 57.5 & 25.5 & 56.3 & 244 & 0.80 \\
    4.84 & -4.80 & 0.60 & 6.4 & 61.5 & 23.4 & 72.6 & 239 & 0.83 \\
    5.06 & -5.00 & 0.80 & 7.0 & 73.0 & 21.5 & 54.0 & 239 & 0.87 \\
    5.30 & -5.20 & 1.00 & 7.4 & 88.1 & 19.1 & 47.0 & 230 & 0.89 \\
    5.60 & -5.45 & 1.30 & 8.0 & 116.9 & 14.9 & 31.6 & 224 & 0.92
    \\ \hline 
  \end{tabular}
\end{table*}
\begin{figure*}
\begin{center}
\includegraphics[width=5.5cm,height=4.3cm]{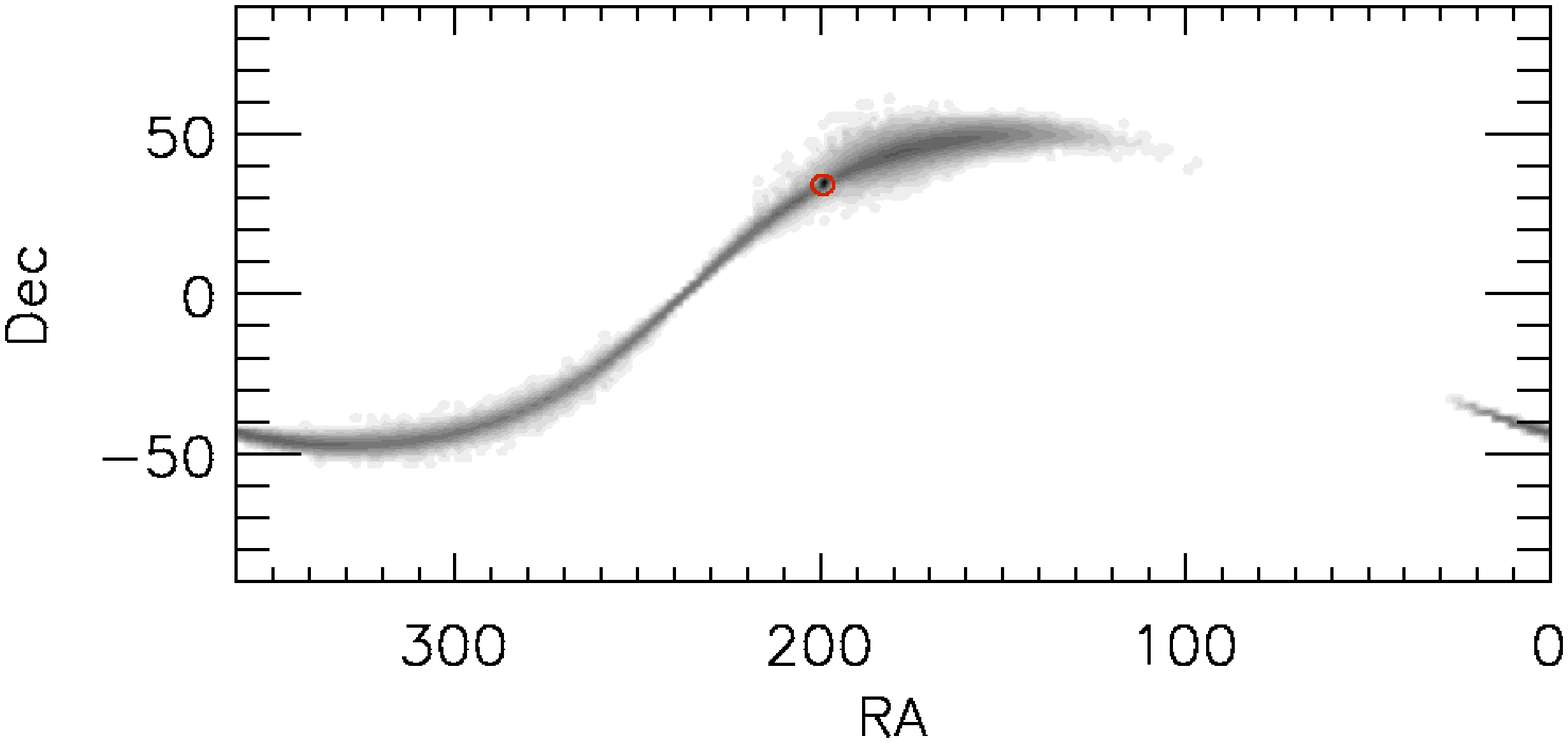}
\includegraphics[width=5.5cm,height=4.3cm]{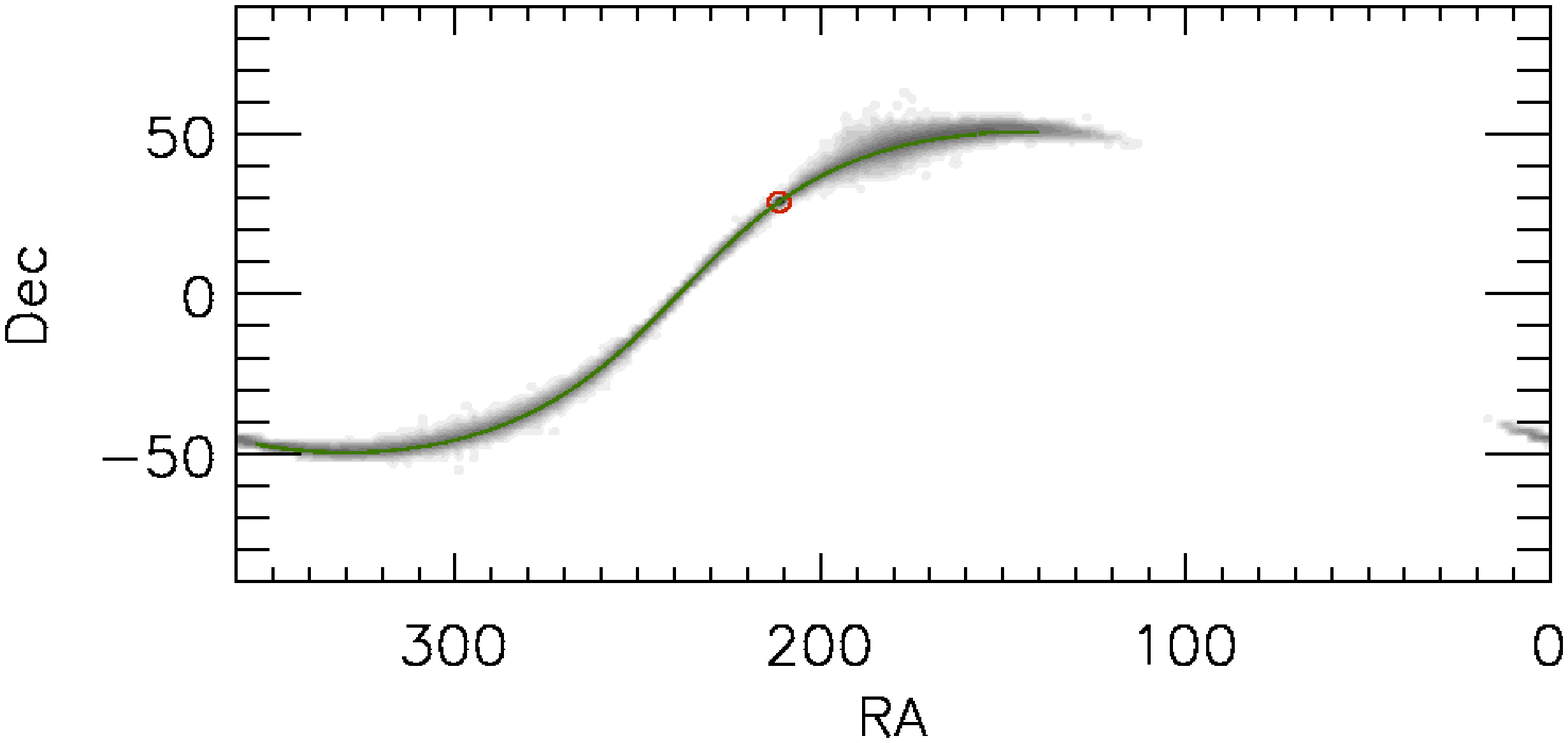}
\includegraphics[width=5.5cm,height=4.3cm]{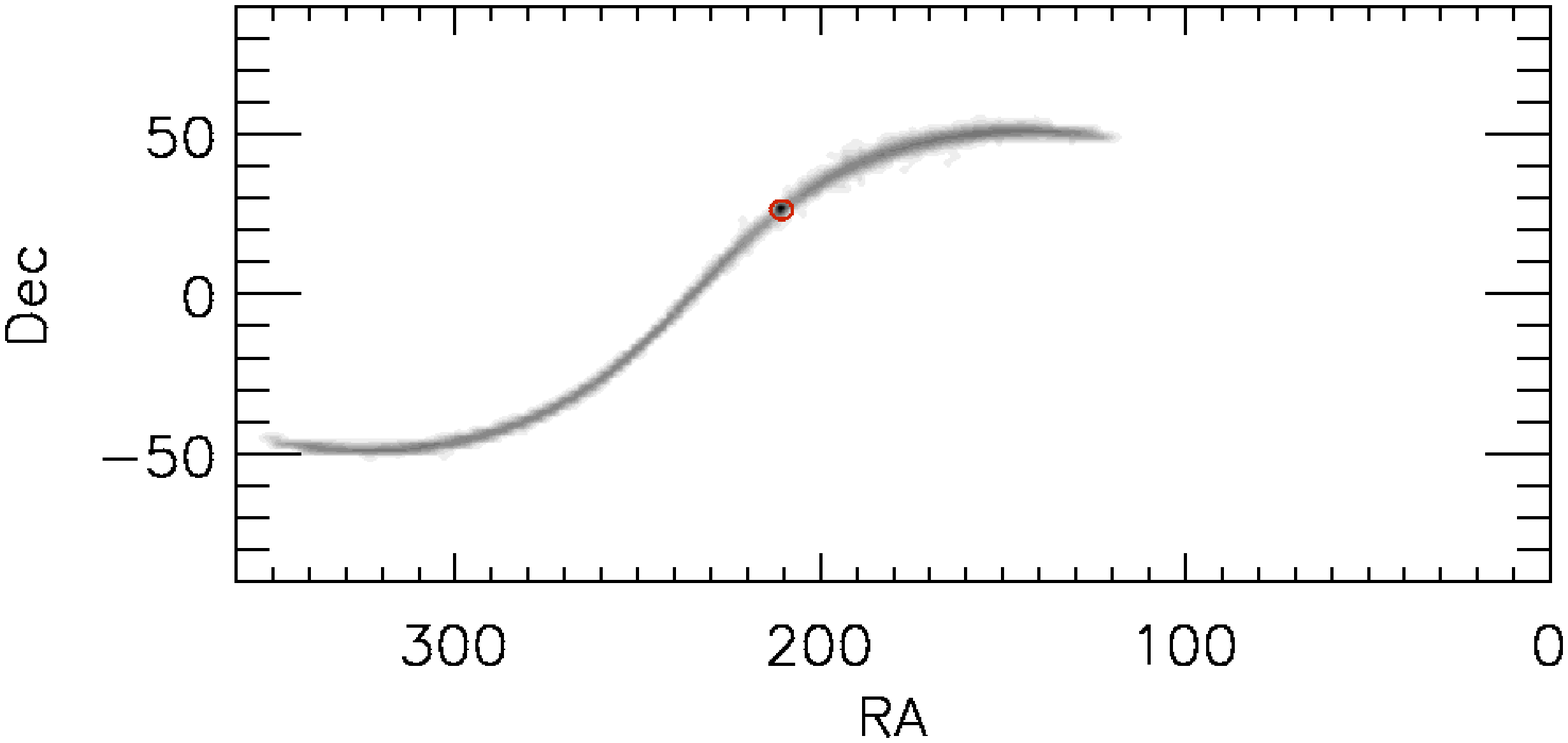}
\includegraphics[width=5.5cm,height=4.3cm]{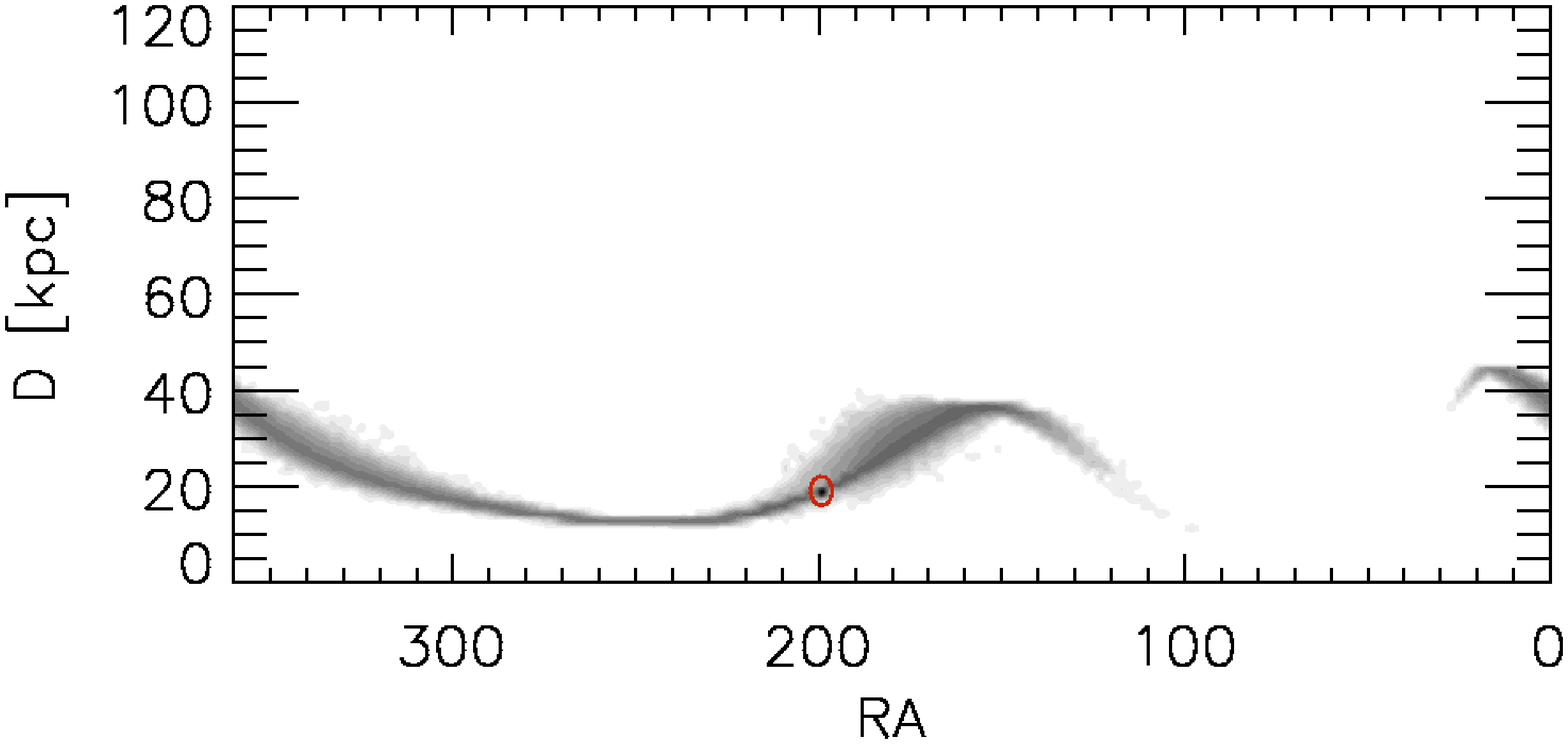}
\includegraphics[width=5.5cm,height=4.3cm]{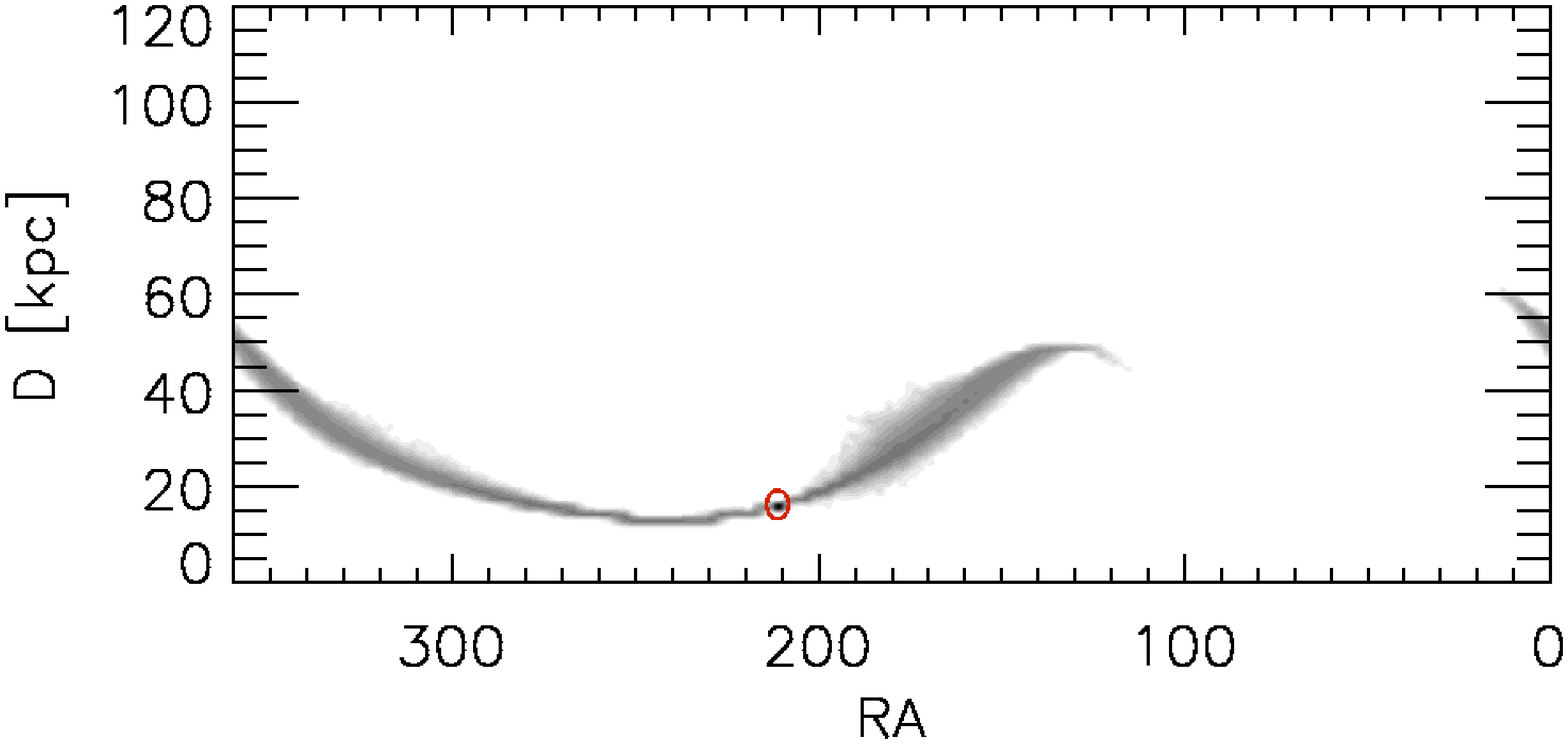}
\includegraphics[width=5.5cm,height=4.3cm]{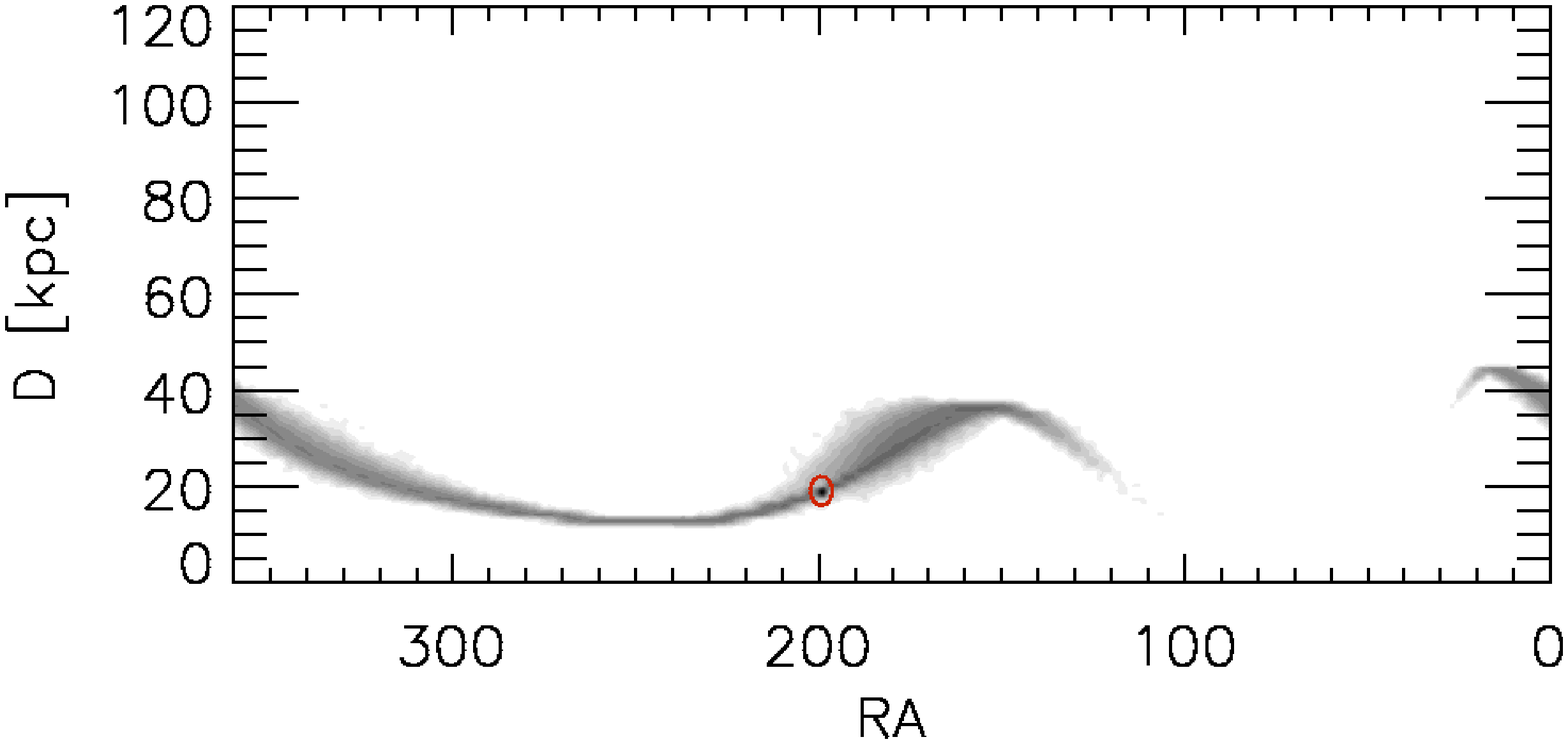}
\includegraphics[width=5.5cm,height=4.3cm]{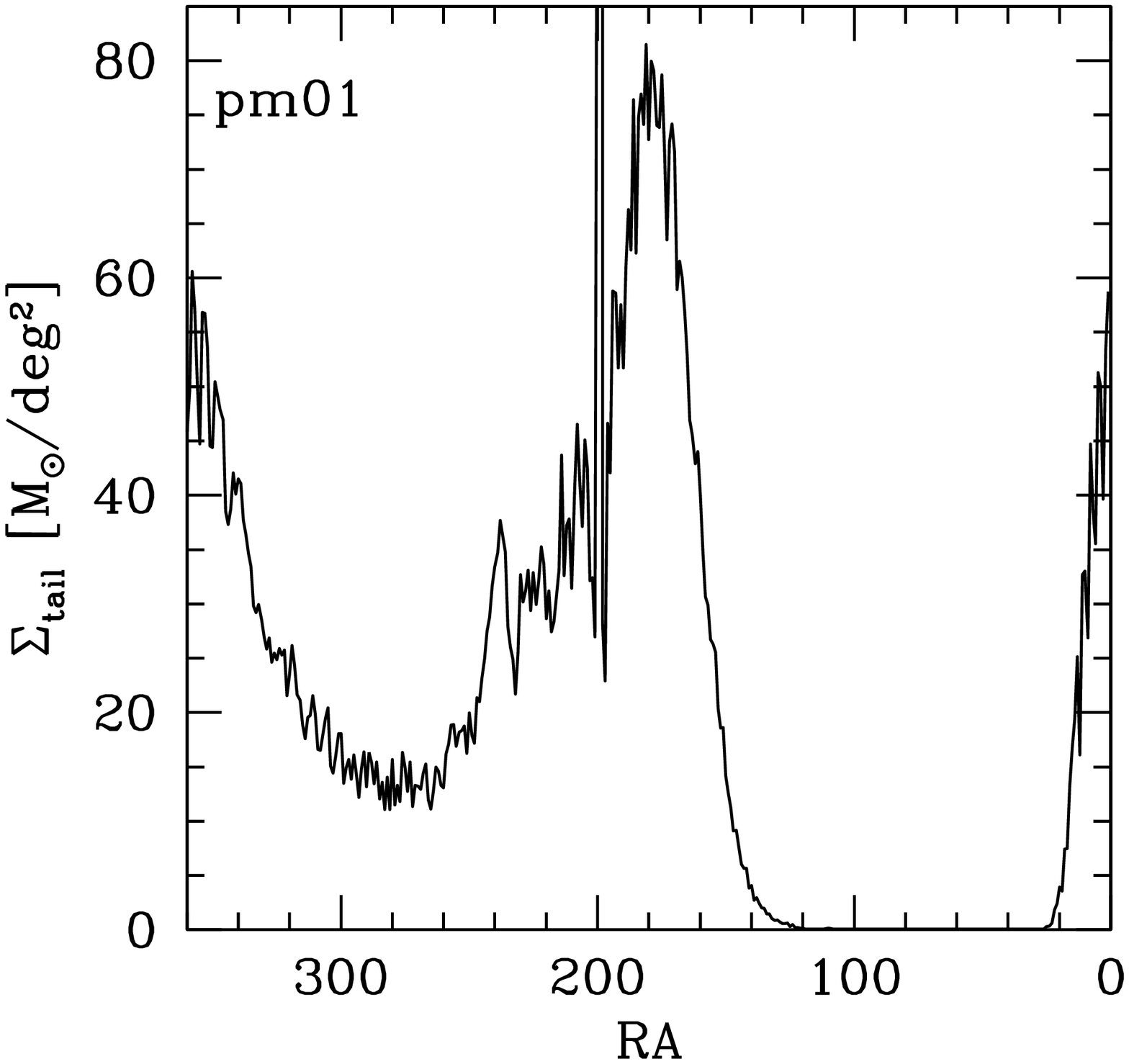}
\includegraphics[width=5.5cm,height=4.3cm]{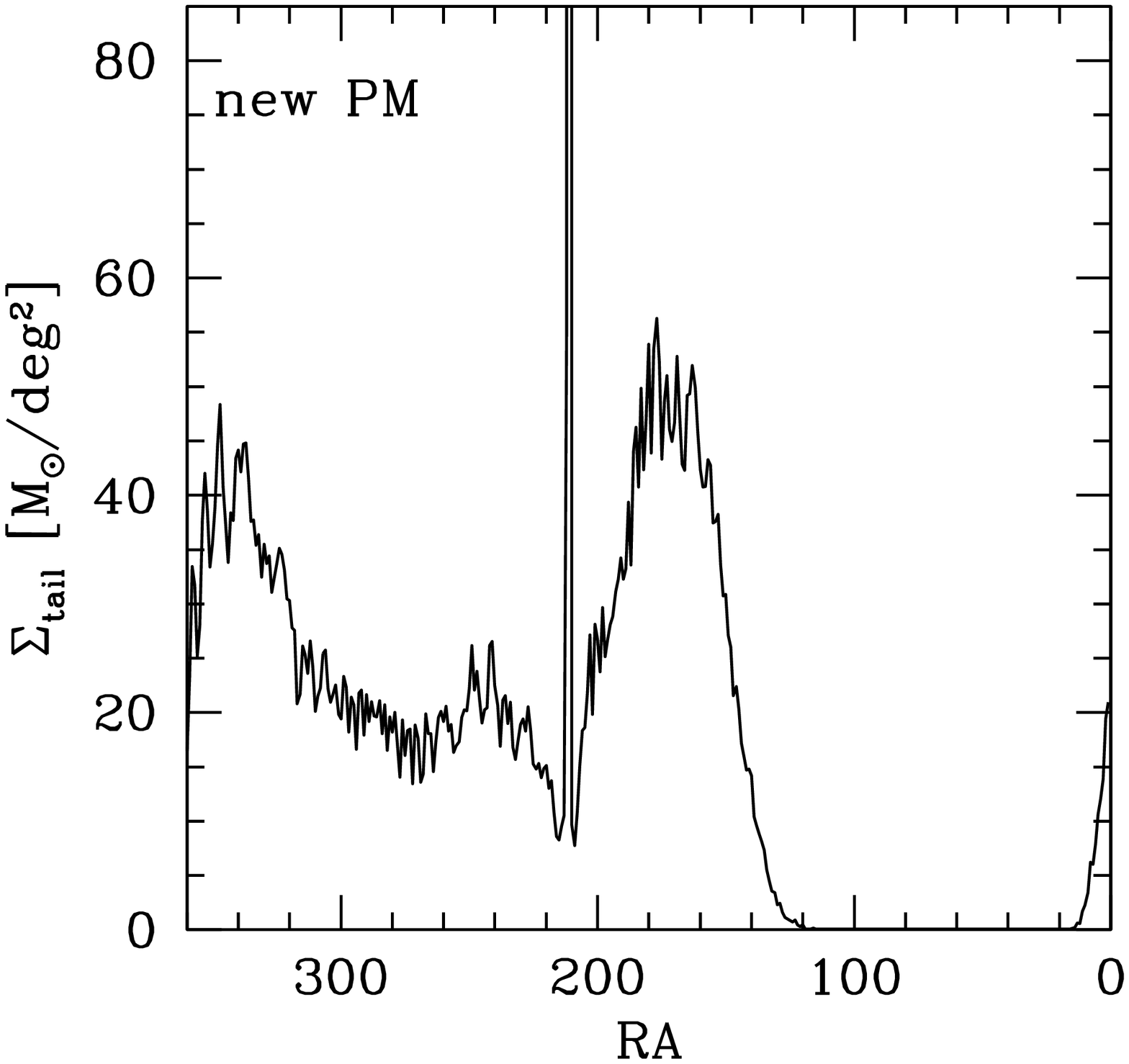}
\includegraphics[width=5.5cm,height=4.3cm]{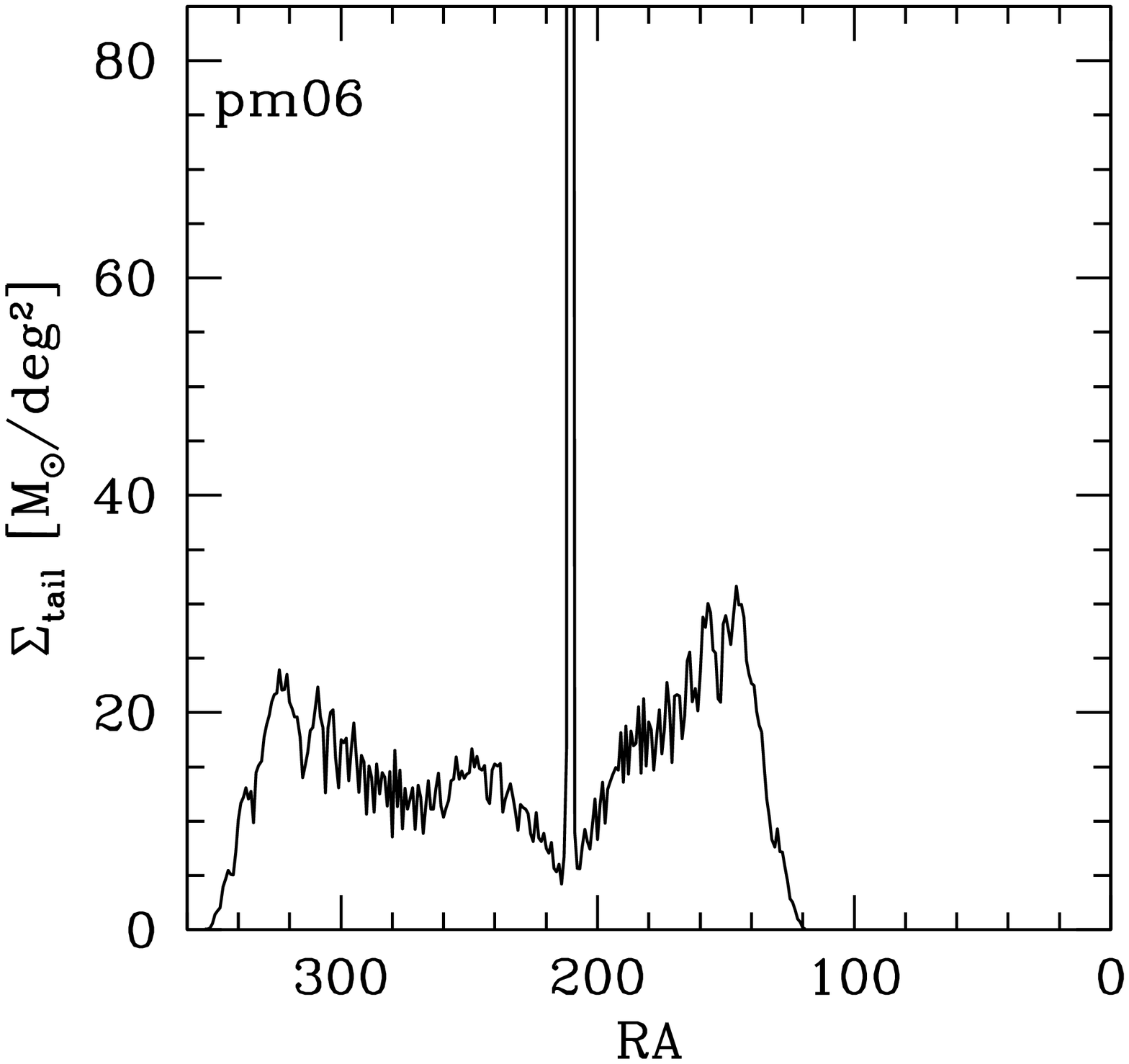}
\end{center}
  \caption{Simulations with different proper motions.  The rows show
    from top to bottom: the projection of the tidal tails in $\alpha$
    and $\delta$, the distance distribution of the tails vs.\ $\alpha$
    and finally the surface density of the tails vs.\ $\alpha$.  The
    columns show first the simulation with the smallest magnitude in
    proper motion, our best fitting model and finally the model with
    the highest magnitude of proper motion.}
  \label{fig:pmdens}
\end{figure*}

In the next two panels of Fig.~\ref{fig:cont}, we show the projection
on the sky of our models in the ML and DB potentials after $10$~Gyr of
evolution.  Both models show faint tidal tails, which match the
general shape of the contours well. However, one important feature of
the data is not reproduced -- the leading and trailing tails are
well-aligned with the proper motion vector. This contrasts with the
data, in which the inner parts of the leading tails are slightly below
the proper motion vector, whilst the inner parts of the trailing tails
are slightly above.  However, the observed proper motions are not well
determined and have large error-bars, so one possibility is that the
proper motion of NGC~5466 should be either larger in right ascension
or smaller in declination than the values given in the literature
\citep[e.g.][]{Di99} to match the observed misalignment.  We confirm
this result by running another simulation with slightly changed proper
motions, namely
\begin{eqnarray}
\mu_{\alpha} \cos \delta & = & -4.7\ {\rm mas\,yr}^{-1} \nonumber\\
\mu_{\delta} & = & 0.42\ {\rm mas\,yr}^{-1}.
\label{eq:revised}
\end{eqnarray}
This orbit gives a perigalacticon of $5.9$~kpc and an apogalacticon of
$57.5$~kpc.  Although the change in proper motion does not make a
significant difference to the mass-loss rate, as shown in the right
panel of Fig.~\ref{fig:mass}, it does improve the match with the
location of the observed tidal streams much better, including the
misalignment. In the lower right panels of Fig.~\ref{fig:cont}, we see
that the inner parts of the leading tails are now slightly below the
old proper motion vector, whilst the inner parts of the trailing tails
are now slightly above.

\subsection{The Tail Densities and Extent}

Fig.~\ref{fig:allsky} shows all-sky views of the tidal tails, together
with density profiles obtained by counting particles. The surface
density of the tidal tails falls off along the innermost tails very
steeply and stays at a very low density of
$20$--$50$~M$_{\odot}$\,deg$^{-2}$ throughout the tails.  These low
densities are very hard to detect, even in surveys like SDSS.
\citet{gr06} found long, almost linear and very tenuous tidal
extensions to NGC~5466 using a matched filter.  Although these
extensions are hard to see in the SDSS data, they do receive some
support from the simulations presented here.  The tails of our model
with the revised proper motion extend over $\sim 100^\circ$ on the sky.
\citet{gr06} claim that the average density of the tails is about
$10$--$20$ stars\ deg$^{-2}$, which is also in good agreement with out
estimate.  Interestingly, at the point where \citet{gr06} start to
loose track of the leading arm, our model is close to its
apogalacticon and the tails are spread out much wider than is the case
close to the cluster.

Although our simulation with the revised proper motion provides a good
representation of the data, it is clearly not unique. In particular,
it is interesting to understand the variety of tidal tail morphology
for NGC~5466, especially as forthcoming deeper photometry will provide
stronger constraints on the modelling. Accordingly, we perform a suite
of {\it Superbox} simulations to investigate how the choice of proper
motion influences the mass-loss and hence the properties of the tidal
tails.  As a constraint, we only used proper motions which are within
the $1\sigma$ error range of the observed value~\citep{Di99} and also
require that the orbital path near the cluster aligns with the tails
found by \citet{Be06a}, i.e., have the same projected orbital path as
our refined set of proper motion.  All-sky views of selected
simulations are shown in Fig.~\ref{fig:pmdens} and show significant
differences in the morphology and the properties of the tails.
Table~\ref{tab:pm} gives the parameters and the results for the entire
suite of simulations.

The number of degrees in right ascension over which the tail is
detectable represents a measure for the length of the tails.  The mean
density of the tails is calculated in the following way.  We examine
one degree in right ascension $\alpha$ and search for the highest
surface density in the tails for each degree in declination $\delta$.
From these values, we compute the average surface density over the
range of right ascension for which the tails are present.  The maximum
density given in the table is computed from the square degree of the
tails with the highest surface density.  Effects of varying distances
are not taken into account.  Table~\ref{tab:pm} shows clearly that the
closer the orbit is to the Galactic centre the more severe is the
mass-loss and the higher is the density in the tails. 

\begin{figure}
\begin{center}
  \includegraphics[height=4.1cm]{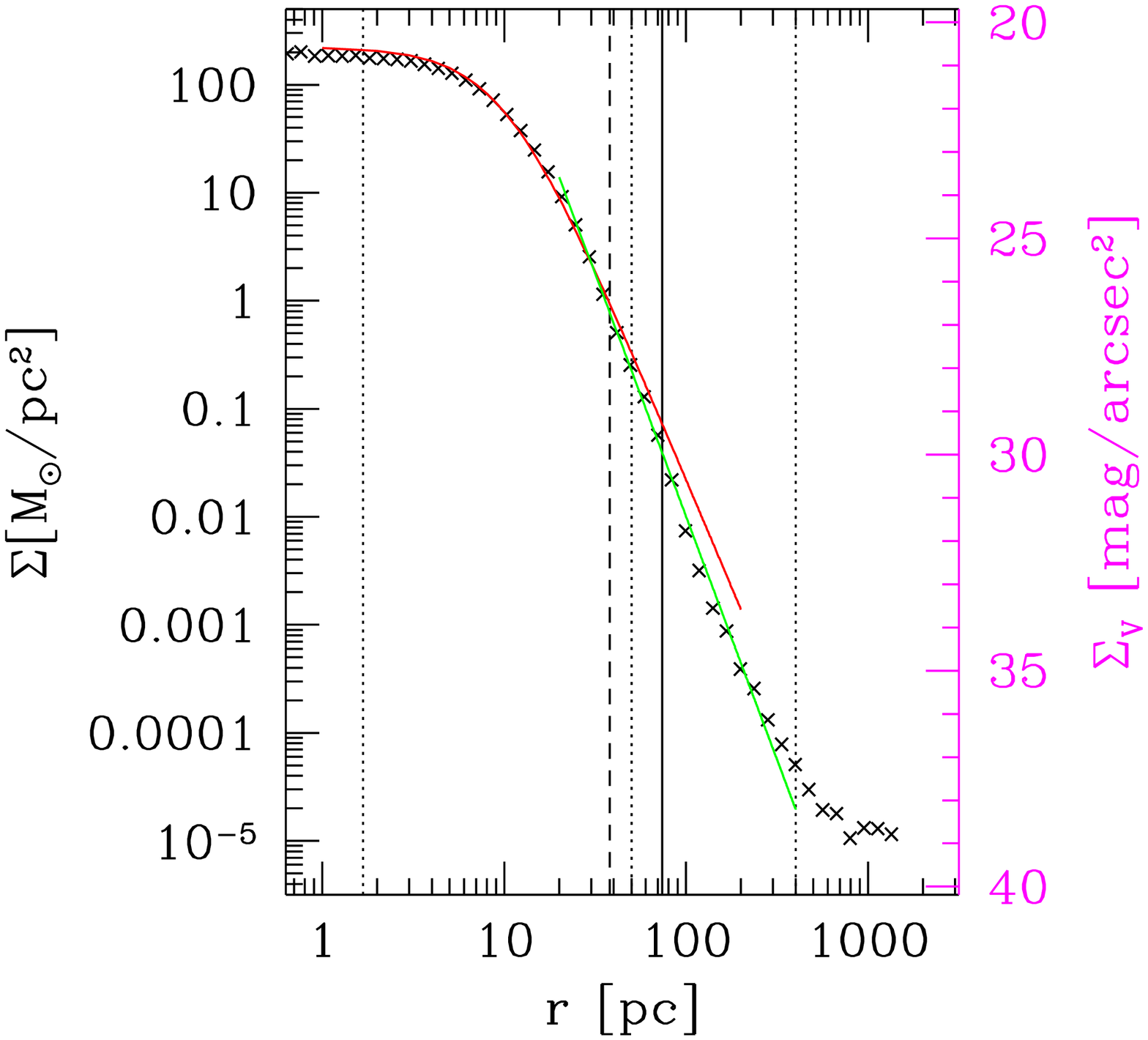}
  \includegraphics[height=4.1cm]{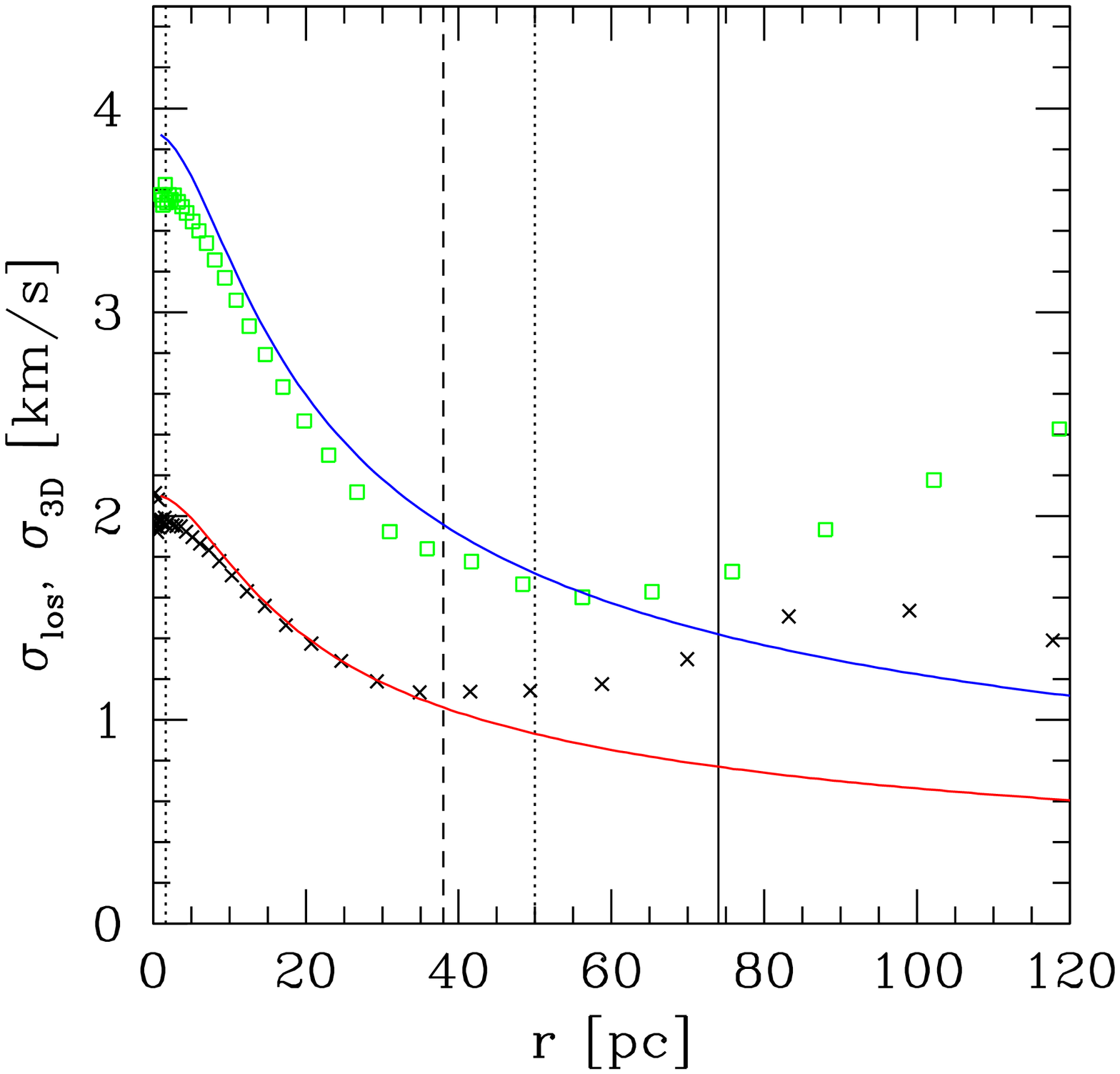}
\end{center}
  \caption{Left: Surface density distribution of our model of the
    NGC~5466 remnant.  Left ordinate shows M$_{\odot}$\,pc$^{-2}$,
    right ordinate shows mag\,arcsec$^{-2}$ using the M/L-ratio from
    literature of $1$.  The inner part is still well fitted by the
    initial Plummer profile, while the outer parts are better fitted
    by a steeper power-law with index $-4.5$.  The deviation is
    visible in the region from tidal radius at perigalacticon to tidal
    radius at apogalacticon.  Vertical lines denote the tidal radius
    now (solid) and at the last perigalacticon (dashed), while dotted
    lines mark the grid boundaries with changes in resolution. Right:
    Velocity dispersion profile.  Green open squares denote the 3D
    velocity dispersion measured in concentric shells around the
    cluster centre, crosses are the line-of-sight velocity dispersion
    measured in concentric rings around the cluster centre.  Curves
    show the profile of the initial model.  Vertical lines are as in
    the left panel.}
  \label{fig:internal}
\end{figure}

\subsection{The Remnant NGC~5466}

Let us consider the internal properties of our remnant cluster in a
representative simulation.  We choose the one which uses the ML
potential and the revised proper motions, shown in the lower right
panels of Fig.~\ref{fig:cont}.  For this simulation,
Fig.~\ref{fig:internal} shows the surface density and velocity
dispersion profiles of the final cluster.  Adopting the data from
\citet{Ha96} (updated values from 2003), the cluster has a central
surface brightness of $21.28$~mag\,arcsec$^{-2}$. This is in good
agreement with our simulation, for which the central surface
brightness is $20.6$~mag\,arcsec$^{-2}$, especially taking into
account that our particle-mesh code neglects internal evolution, which
would lead to higher densities in the core.  Also, the half-light
radius in our simulation is $10$~pc and corresponds well with the
observed values of $10.4$~pc \citep{Ha96} and $13.1$~pc \citep{Pr91}.
The actual tidal radius in our model is $75$~pc \citep[using the
  Jacobi limit as given in][]{Bi87} and is less than the $158$~pc
stated in \citet{Ha96} ($158$~pc) or $97$ pc in \citet{Le97}, but only
slightly larger than the radius of $61$ pc found by \citet{Pr91}. Note
that, observationally speaking, $r_{\rm tidal}$ is determined by
fitting a King (1966) profile to the surface brightness distribution,
which does not correspond exactly to the theoretical definition.
According to our simulations, the tidal radius at the last
perigalacticon was about $38$~pc.

While the surface density in the inner parts is not much affected by
the mass-loss, the central velocity dispersion is reduced by
approximately $10$~\%.  Also visible is a rise in the line-of-sight
velocity dispersion in the outer parts, which starts already within
the actual tidal radius.  This is due to the fact that all
line-of-sight measurements are contaminated by unbound stars streaming
in front or behind the star cluster.  While they do not affect the
central values because of their low number, their effect is easily
measurable in the outer parts where the densities of the bound stars
are much lower.

\section{Conclusions}
\label{sec:conclus}

We have presented numerical simulations of the formation and evolution
of the tidal tails of the globular cluster NGC~5466. We used direct
N-body codes to argue that the evolution of the cluster is dominated
by external effects rather than internal relaxation, and then
grid-based codes to trace the faint tidal tails.  This novel, hybrid
approach is well-suited to map out the detailed morphology of the
low-density tails of NGC~5466.

Naively, we might expect that a low mass cluster with observed and
very lengthy tails on a disc crossing orbit would not be able to
survive for too much longer.  However, simulations by \citet{De04} have
already shown that the disrupting globular cluster Pal~5 has survived
for at least many Gyr in a tidally-dominated and out-of-equilibrium
state, although Pal 5 probably will be destroyed at the next
disc crossing. Here, we have demonstrated that a progenitor cluster of
NGC~5466, which is quite similar to the present cluster, could survive
substantially longer, for at least a few Hubble times, with its
extensive but tenuous tidal tails gradually wrapping around the whole
Galaxy.

The evolution of NGC~5466 is mainly driven by tidal shocks at each
perigalacticon/disc crossing combination.  Although not entirely
negligible, internal effects (two-body relaxation and evaporation of
stars driven by post-core collapsed processes \citep{Le95}) play a
much less important role in the mass-loss.  It is this property which
allows us to study the tidal tails using grid-based codes rather than
the more cumbersome direct N-body codes.  If the observationally
determined mass-to-light ratio of $\sim 1$ is correct, then the
initial mass of NGC~5466 is $\sim 7 \times 10^{4}$~M$_{\odot}$.  By
initial mass, we do not mean the embedded mass of the star cluster at
its formation inside a gas-cloud.  If the star formation efficiency is
low, a star cluster can lose about $\sim 70$~per cent of its initial
mass in stars when the gas gets blown out by high velocity winds or
supernovae explosions.  The rapid stellar evolution of high mass stars
then adds another extreme mass-loss of $\sim 20$~per cent in the first
few tens of Myr.  After this initial phase of rapid evolution, the
cluster reaches a quasi-equilibrium.  This is the starting point of
our simulations and therefore our initial mass refers to this point in
time. 

Our numerical simulations reproduce the observational results of both
groups who have recently studied the tidal tails NGC~5466 with SDSS
data. Mapping out the tails close to the globular cluster,
\citet{Be06a} found that the leading tail emerges from the side
pointing towards the Galactic Centre and returns to the orbital path
from outside, while the trailing tail emerges from the side opposite
to the Galactic Centre and returns to the orbital path from within.
With our simulations, we showed that the proper motion of the globular
cluster has to be smaller in declination and/or larger in right
ascension than reported by~\citet{Di99} to account for the position of
the tidal tails.  We propose a new set of proper motions,
$\mu_{\alpha} \cos \delta = -4.7\ {\rm mas\,yr}^{-1} \mu_{\delta} =
0.42\ {\rm mas\,yr}^{-1}$ for which the tail morphology is correctly
reproduced.  This differs from the observationally determined one by
$-0.05$ and $-0.38$~mas\,yr$^{-1}$ respectively.  These changes are
within the error margins of the observed proper motion ($\pm
0.82$~mas\,yr$^{-1}$).

The surface density of the tidal tails falls off along the innermost
tails very steeply and stays at a very low density of
$20$--$50$~M$_{\odot}$\,deg$^{-2}$ throughout the tails.  These low
densities are very hard to detect, even in surveys like SDSS.
\citet{gr06} found long, almost linear and very tenuous tidal
extensions to NGC~5466 using a matched filter approach. Their work is
supported by the simulations in this paper, which show that the very
long ($\gta 100^\circ$), faint tidal tails are expected. The tails in
our simulation have roughly the same surface density as found by
\citet{gr06}.

In the future, deeper photometry, radial velocities and -- thanks to
the GAIA and SIM satellites -- proper motions of individual stars in
the tidal tails may become available.  Mapping out the structure of
the tails of globular clusters and dwarf galaxies will then provide
powerful constraints on the Galactic potential. This work, together
with the observational papers of \citet{Be06a} and \citet{gr06}, has
shown that NGC~5466 is a prime target for such studies of cold
streams. Its tidal tails, though faint, are the longest so far claimed
for any Milky Way globular cluster.
\\ 

\noindent {\bf Acknowledgements:} MF and VB are funded by PPARC.
MIW acknowledges support from a Royal Society University Research
Fellowship.  We thank W. Dehnen for providing his Galactic potential
code and R.Spurzem and H.M. Lee for useful comments.  The direct
N-body simulations were performed on the GRACE supercomputer at
ARI-ZAH Heidelberg funded by Volkswagen Stiftung I/80 041-043 and the
State of Baden-W\"{u}rttemberg, using GRAPE hardware.

\label{lastpage}

\end{document}